\documentclass[prl,superscriptaddress,notitlepage,showpacs,twocolumn]{revtex4-2}
\usepackage{amssymb,color,xcolor,float,xspace,mathtools,array,tikz,txfonts}
\usepackage[bottom]{footmisc}
\usepackage{graphicx,footnote}
\usepackage{amsmath}
\usepackage{mathrsfs}
\usepackage{wrapfig}
\newcommand{\om}{{\omega}}
\usepackage{hyperref}

\begin{document}
\title{Thermalization and Mpemba-like patterns in effective temperature dynamics of strongly coupled dissipative quantum chaotic systems}
\author{Xuanhua Wang}
\email{Corresponding author: wangxh@fyust.edu.cn}
\affiliation{Fuyao University of Science and Technology, Fuzhou, Fujian 350122, China}

\author{Jie Su}
\thanks{Co-first author.}
\affiliation{Center for Theoretical Interdisciplinary Sciences, Wenzhou Institute, University of Chinese Academy of Sciences, Wenzhou, Zhejiang 325001, China}

\author{Jin Wang}
\email{Corresponding author: jin.wang.1@stonybrook.edu}
\affiliation{Center for Theoretical Interdisciplinary Sciences, Wenzhou Institute, University of Chinese Academy of Sciences, Wenzhou, Zhejiang 325001, China}
\affiliation{Department of Chemistry, Stony Brook University, Stony Brook, New York 11794, USA}
\affiliation{Department of Physics and Astronomy, Stony Brook University, Stony Brook, New York 11794, USA}

\begin{abstract}
Anomalous thermalization, particularly the crossings of temperature trajectories from different initial states termed Mpemba crossings (MPCs), have intrigued scientists for decades. While recent studies in quantum systems suggest that initial conditions play a decisive role in its emergence, they offer limited insight into MPCs in complex, highly nonequilibrium systems. In this study, we investigate temperature dynamics in the strongly coupled, quantum chaotic Sachdev-Ye-Kitaev (SYK) model, which is dual to the low-energy dynamics of 2D dilaton gravity. Our findings reveal a dynamically driven nonequilibrium mechanism underlying MPCs during rapid thermalization, with implications for gravitational systems. We explore quench dynamics in SYK systems under three conditions: coupling to a single SYK thermal bath, coupling to two thermal baths at different temperatures, and dissipative SYKs modeled by the Lindblad equation. We find that strong system-bath coupling induces oscillating effective temperatures and trajectory crossings in transient states due to nonequilibrium statistics, phenomena absent in quasi-static thermodynamics and Lindbladian SYKs. These MPCs highlight a unique feature of anomalous thermalization of strongly coupled quantum chaotic systems driven far from equilibrium. Besides, the results also provide qualitative insights into the nonequilibrium thermodynamics of black holes strongly interacting with their environment, such as primordial black holes in the early universe.
\end{abstract}

\maketitle
%
%
\section{Introduction.}
\begin{figure}[t]
    \centering
    \includegraphics[width=.9\columnwidth]{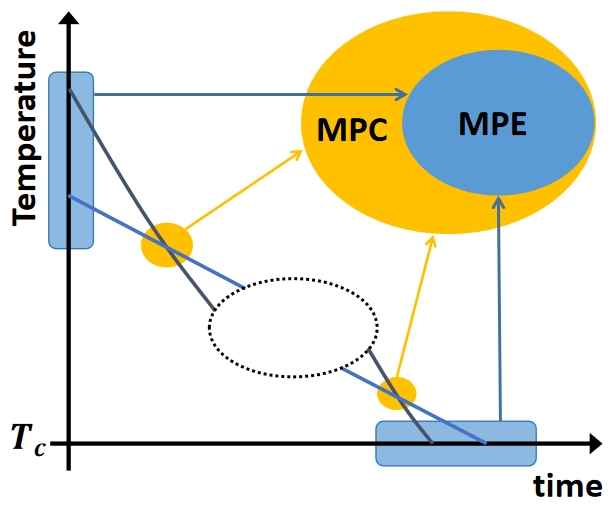}
    \caption{Diagrammatic illustration of MPEs and MPCs. The slanted black and blue curves are examples of temperature trajectories from two initial states. The region inside the dotted circle represents other unknown dynamics. The MPE is usually defined as the order exchange of the initial and the final states in the blue shaded regions along the x- and y-axes. The final state is at a certain critical temperature $T_c$, corresponding to a phase transition temperature such as the icing point or the temperature of the surrounding heat bath. The MPC refers to the crossing of the temperature trajectories during the dynamic process, indicated by the orange circled regions at the intersections of the two curves. MPCs can be viewed as an generalization of the MPE.}
    \label{fig:MPC}
\end{figure}

Relaxation dynamics have unveiled a plethora of phenomena that lie beyond the framework of equilibrium thermodynamics. A prime example is the Mpemba effect (MPE), which refers to the surprising observation that hot milk can freeze faster than cold milk when both are subjected to the same conditions \cite{mpemba1969cool}. This effect serves as a striking illustration of counterintuitive behaviors in thermalization. 
Although the reproducibility of the original MPE remains a topic of debate \cite{burridge2016questioning,jin2015mechanisms}, its hallmark feature--the crossing of temperature trajectories, known as the Mpemba crossing (MPC) [see Figure \ref{fig:MPC}]--has found manifestation in broader systems \cite{chalas2024multiple,klich2019mpemba,wang2024mpemba}. MPCs challenge the conventional quasi-static thermodynamics, such as the Stefan-Boltzmann law since the cooling rate according to the conventional thermodynamics is a state function and no temperature crossings are allowed. 
In recent years, diverse manifestations of MPCs and other Mpemba-like effects have been observed in an increasing range of media, including water \cite{jeng2006mpemba}, granular fluids \cite{lasanta2017hotter,biswas2020mpemba}, molecular gases \cite{megias2022thermal,santos2020mpemba}, and ion-trap quantum computers \cite{shapira2024mpemba,joshi2024observing,zhang2025observation}. The growing body of evidence calls for a deeper understanding of the criteria for the MPC emergence and the underlying mechanisms that govern it \cite{lu2017nonequilibrium}. 

Recent theoretical advances based on models of quantum spin chains have shed light on its potential mechanisms  \cite{wang2024mpemba,Moroder24,rylands2024microscopic,chatterjee2023quantum,carollo2021exponentially,murciano2024entanglement,yamashika2024entanglement,liu2024symmetry,chatterjee2024multiple,nava2024mpemba}. However, the analyses so far have mainly focused on simple quantum integrable models, such as the quantum dot and the minimal Kitaev model, where MPC emergence is found to be dictated by initial conditions and settings of the baths. This is inconsistent with experimental findings which indicate that nonequilibrium dynamics is the key. Furthermore, in models that better resemble complex classical medium such as quantum chaotic models, information about a finely tuned initial state will be rapidly scrambled and lost, likely invalidating the initial-condition dependency \cite{maldacena2016bound,touil2021information,ares2025quantum,liu2024symmetry,turkeshi2024quantum,qian2024intrinsic}. 
Whether MPCs can arise in strongly interacting chaotic systems and extend to larger scales through the same initial-condition mechanism remains an open question.

The emergence of MPCs in quantum systems has far-reaching implications for our understanding of fundamental physics and cosmology. Not only the quantum spin models, black holes--like the liquids in Mpemba's experiment--can also exchange energy with the surroundings through Hawking radiation and attain equilibrium in certain bounded or contracting universes such as anti-de Sitter space (AdS) \cite{hawking1975particle}. Besides the qualitative analogy, contemporary interpretations based on gauge/gravity duality provide a precise correspondence between the thermodynamics of black holes and that of strongly interacting quantum matter \cite{maldacena1999large}. The deep connection inspires questions about black hole thermalization: whether they can exhibit anomalous dynamics like quantum spin chains. This question is crucial in the early universe where hot and dense matter permeates \cite{dodelson2020modern,gorbunov2011introduction}. Since static black holes behave as ideal black-body radiators that allow no thermalization anomalies, corrections to this depiction in a dynamical process would be required for temperature crossings to occur, leading to an extra factor that could influence the big-scale structure and early temperature fluctuations. Given the turbulent thermodynamic history of the universe during its early epoch and persistent discrepancies between observations and theoretical models \cite{aghanim2020planck,di2021realm,riess2022comprehensive,bertone2018history}, questions about thermalization anomalies in black holes seem timely and compelling. 

Though direct numerical computation of open black hole dynamics including Hawking radiation is highly demanding, the Sachdev-Ye-Kitaev (SYK) model, as a solvable toy model of gravitation that also exhibits quantum chaos, provides an optimal entry point to shed light on both issues 
\cite{sachdev1993gapless,Kitaev,maldacena2016remarks,gu2017local}. Originally conceived to study many-body chaos in spin-S Heisenberg models \cite{sachdev1993gapless}, the SYK model describes the dynamics of Majorana fermions with long-range random interactions. 
It has garnered significant attention since the discovery of its ability to describe the dynamics of a gravitational theory theory in AdS named Jackiw-Teitelboim (JT) gravity \cite{Kitaev}. Apart from the common Schwarzian dynamics, the two theories also show similar properties in their thermodynamic quantities such as the entropy and the rate of information scrambling, which is quantified by the Lyapunov exponent, providing a great platform to study both black holes and strongly-correlated chaotic systems \cite{scaffidi2019chaos,maldacena2016bound,gu2017local,polchinski2016spectrum,shenker2014multiple,touil2021information}. 
Besides isolated SYKs, coupled and dissipative SYK models, especially their real-time Hamiltonian and quench dynamics, have also attracted extensive research recently \cite{chen2017tunable,garcia2021euclidean,garcia2023keldysh,garcia2022dominance,zhang2019evaporation,zanoci2022energy,eberlein2017quantum,almheiri2024universal,maldacena2021syk,cheipesh2021quantum,bhattacharya2019quantum,sa2022lindbladian,kawabata2023dynamical,wang2024entanglement}. 
Though the exactness of the gravity duality is more obscure in open systems, many essential features of the model persist. For instance, the dissipative SYK models still exhibit quantum chaos, albeit with reduced Lyapunov exponents depending on the system-environment coupling \cite{chen2017tunable,garcia2024lyapunov}, and dynamical transitions in coupled SYK models reveal transitions between traversable wormholes and black holes in gravity \cite{garcia2023keldysh,kawabata2023dynamical,garcia2021euclidean,garcia2022dominance,li2021free}. 

In this study, we investigate the thermalization of SYK models coupled to baths and discuss its implications for black hole thermodynamics. 
For SYK models that slowly thermalize, the temperature evolution exhibits smooth asymptotic approach towards the steady-state temperature without crossings. In the regime of rapid cooling, however, we observe not only the emergence of MPCs in the system temperatures, but also the transient overcooled and overheated phases that oscillate around the steady-state temperature once the system-bath coupling exceeds a certain threshold as shown in Fig.~\ref{fig:Fig_2}. The dynamical MPCs are caused by states following different thermalization paths away from thermal states. This behavior is only present in the exact calculations of the system dynamics coupled to SYK baths and absent in the phenomenological description of the dissipative SYK model in the Lindbladian formalism due to the crude modeling of the interaction with bath. 
These findings have implications for dynamic black holes undergoing rapid evolutions, such as the primordial black holes in the early universe where dense external matter of different temperatures interacts strongly with them. The potential MPCs during their fast relaxation can lead to Hawking radiation deviated from the thermal spectrum. Numerical simulations incorporating both classical black hole dynamics in dissipative systems and their quantum
Hawking radiation or direct observational evidences will be needed to confirm these implications. 

\section{Results}
\subsection{Theoretical background}

\subsubsection{MPC mechanisms in existing research and limitations} 
Many theoretical analyses on MPCs have focused on Markovian models, where properties of measured observables and initial conditions dictate the occurrence of these anomalies. It is worth noticing that the rapidity of the dynamics, which is one of the keys to the emergence of the MPE in experiments, is not the determining factor in these models. In a quantum system, a state further from the steady state—based on measures such as entanglement asymmetry and trace distance—may not be the state furthest from equilibrium during thermalization and cause the emergence of MPEs \cite{ares2023entanglement,murciano2024entanglement,yamashika2024entanglement,liu2024symmetry,chalas2024multiple}. For an N-level quantum system weakly interacting with its environment, one can prepare the initial state of the system as the fast-decaying mode in the Liouvillian spectrum. Specifically, for a system whose Liouvillian equation of motion for its linearized density matrix $\Vec{\rho}$ approximately given by $\frac{d \vec\rho(t)}{dt}=\mathscr{L}_{T_b} \vec\rho_0$, one can always pick the particular initial state $\rho_{fast}$ such that its inner product with the slowest decaying eigen mode of the superoperator $\mathscr{L}_{T_b}$ vanishes. 
However, whether such arguments can be extended to more complex quantum systems such as those exhibiting quantum chaos, where the Hamiltonians behave as random matrices and the specialty of a finely tuned initial state is lost due to the randomness in the ensembles of Hamiltonians and the chaotic dynamics, remains unclear \cite{maldacena2016bound,touil2021information}. 
Quantum chaos, defined as the behavior of a quantum system whose classical limit is chaotic \cite{maldacena2016bound,scaffidi2019chaos,casati2022quantum}, provides a crucial link between simple quantum dynamics and complex classical behavior. In this study, we will investigate the temperature dynamics in quantum chaotic systems.

\subsubsection{SYK models in thermal baths}
We consider SYK quench dynamics in three different setups: an SYK model coupled with an SYK bath, an SYK model couples with two different baths, and the Lindbladian SYK. We present the formulation of the first case in the main text, and details of the latter two cases are provided in the Supplemental Material (SM) \cite{suppl}. We consider a q-body interacting SYK$_q$ model in a thermal bath, consisting of $N$ Majorana fermions in (0+1) dimensions with random $q$-fermion interactions. The bath is an SYK$_q$ model governed by the same Hamiltonian with $N^2$ Majorana modes. In the large-$N$ and a weak system-bath coupling limit, the model exhibits maximal chaos for $q>2$, saturating the bound on the quantum Lyapunov exponent \cite{maldacena2016bound,polchinski2016spectrum,shenker2014multiple}. For simplicity, we focus on $q=4$ case and the total Hamiltonian is given by
\begin{align}
H=H_{\text{SYK}}[J_{i_1i_2i_3i_4},\chi]&+H_{\text{SYK}}[\tilde{J}_{i_1i_2i_3i_4},\psi] \nonumber\\
&+\sum_{ai_1i_2...i_n}\frac{V_{ai_1i_2...i_n}}{n!}\chi_a\psi_{i_1}\psi_{i_2}\dots\psi_{i_n}.\label{eq:H}
\end{align}
where $\chi_i$ are the Majorana fermions of the system on site $i=1... N$ obeying $\{\chi_i,\chi_j\}=\delta_{ij}$ and $\psi_i$ represent the bath operators on site $i=1... N^2$. $H_{\text{SYK}}[J_{i_1i_2i_3i_4},\chi]$ and $H_{\text{SYK}}[\tilde J_{i_1i_2i_3i_4},\psi]$ are the standard SYK$_4$ Hamiltonians for the system $\chi$ and the bath $\psi$, respectively. They are of the same form given by:
\begin{align}
H_{\text{SYK}}[J_{i_1i_2i_3i_4},\chi]=\sum_{i_1i_2i_3i_4}\frac{J_{i_1i_2i_3i_4}}{4!}\chi_{i_1}\chi_{i_2}\chi_{i_3}\chi_{i_4}\,.
\label{1}
\end{align}
The model can be understood as an ensemble theory with the interaction strengths chosen from the random Gaussian distribution with mean values given by:
\begin{align}
&\overline{J_{i_1i_2i_3i_4}}=0,\ \ \ \ \ \ \overline{\tilde{J}_{i_1i_2i_3i_4}}=0,\ \ \ \ \ \ \overline{V_{ai_1i_2...i_n}}=0,\\
&\overline{J_{i_1i_2i_3i_4}^2}=\frac{3!J^2}{N^3},\ \ \ \overline{\tilde{J}_{i_1i_2i_3i_4}^2}=\frac{3!J^2}{N^6},\ \ \ \overline{V_{ai_1i_2...i_n}^2}=\frac{n!V^2}{N^{2n}}.
\end{align}
The randomness of the SYK Hamiltonian eliminates the significance of specific initial states in causing faster decay rates, as seen in spin models. The system dynamics and temperature will be discussed in the Method section.

\begin{figure}
    \centering
    \includegraphics[width=1.0\columnwidth]{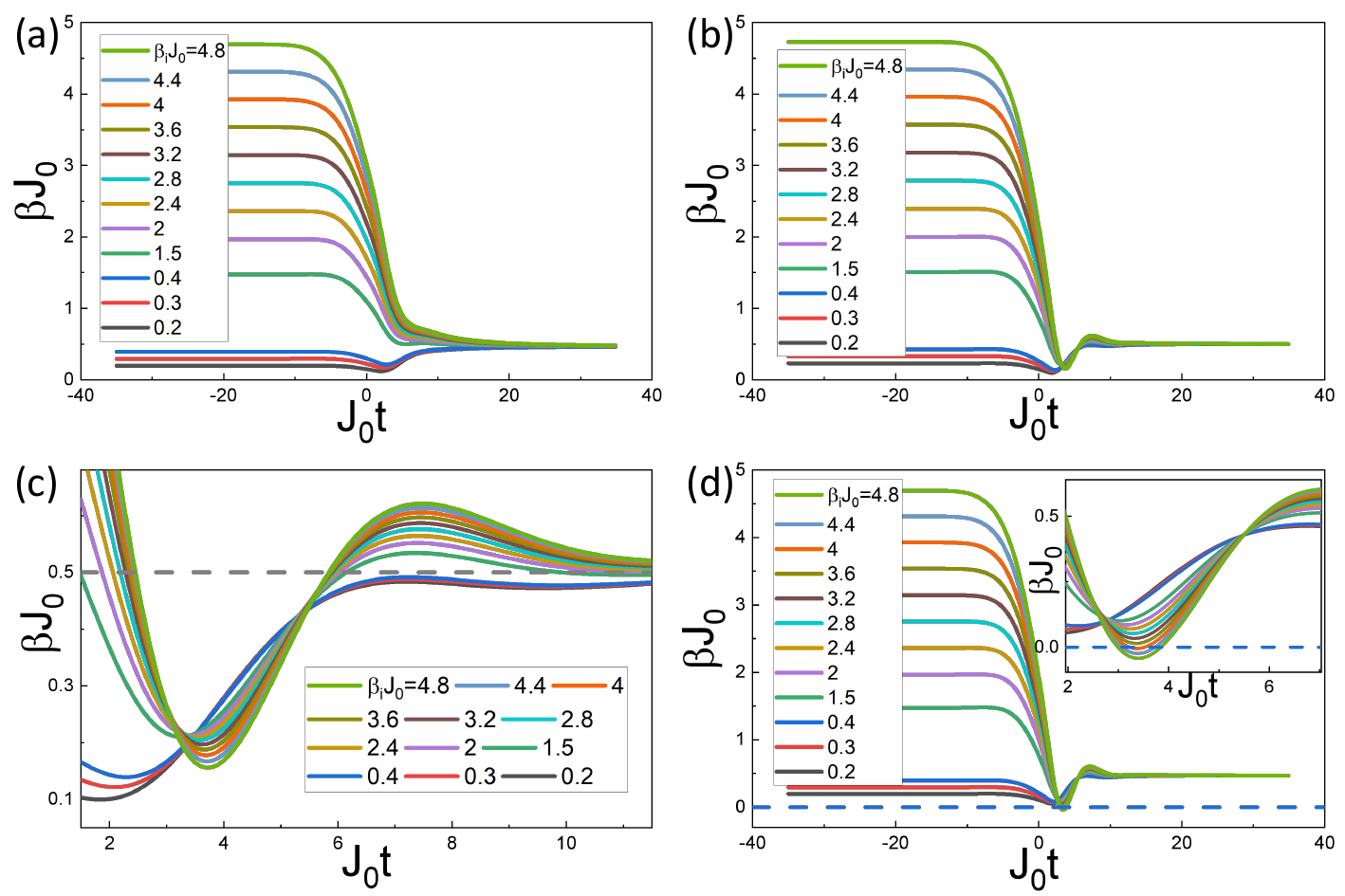}
    \caption{The evolution of inverse temperature $\beta$ vs time $t$. The coupling between the system and the bath is set to be (a) $V=0.3J_0$, (b) $V=0.525J_0$, (c) $V=0.525J_0$ with the zoomed-in view of (b), where the dashed horizontal line indicating the bath temperature, and (d) $V=0.55J_0$ with an inset the zoomed-in view. For weak system-bath couplings, the thermalization processes behave as expected. The anomalous effects only exist for strong couplings. For all, $n=3,\,  \beta_{\mathrm{bath}} J_0=0.5$, and $J_0$ is the reference coupling strength related to the SYK coupling $J$ by $J_0=2J$.}
    \label{fig:Fig_2}
\end{figure}

\subsection{Temperature Oscillations and MPCs}
\subsubsection{Thermalization of SYK models coupled to a bath}
In SYK models, it is nontrivial to accurately quantify the temperature of a system especially during nonequilibrium processes. The most widely used inverse of the effective temperature is given by the ratio between the Keldysh and retarded Green's functions at low frequency \cite{eberlein2017quantum}, namely, 
\begin{gather}
    \beta(t)=\left.\frac{2 \cdot \mathrm{Im}(G_{\chi,K}(\omega,t))}{\omega \cdot \mathrm{Im}(G_{\chi,R}(\omega,t))}\right|_{\omega\rightarrow 0}\,.
\end{gather} 
The discussion on the quantification of effective temperature and the results using an alternative definition of effective temperature are provided in the SM.

We study the effective temperatures of the SYK model during thermalization when coupled to an SYK thermal bath. Initially, we prepare the system to be certain thermal states without coupling to the bath. At time $t=0$, we couple the system to the thermal bath and watch the temperature evolution of the system. For weak couplings to the bath, the cooling dynamics of the system align with the predictions of quasi-static analysis and feature a smooth cooling or heating process [see Figure~\ref{fig:Fig_2}(a)]. In fact, for slow thermalization, the inverse temperature of a coupled SYK model at late times can be approximated by the smooth function $\beta_{\mathrm{eff}}(t)=\beta_f +\alpha \exp (-\Gamma t)$, where $\beta_f$ is the temperature at the final state, $\alpha$ is related to the temperature difference between the initial and the final states, and $\Gamma$ is the thermalization rate independent of initial states \cite{eberlein2017quantum}. This equation is in accordance with quasi-equilibrium thermodynamics and predicts no dynamical anomalies. The dip near $t=0$ is caused by the energy input when coupling is turned on. Notably, when the coupling strength increases, the temperature starts to decay in the manner beyond the quasi-equilibrium description, highlighted by temperature oscillations and the emergence of MPCs. Such phenomena arise due to the system's strong out-of-equilibrium state, induced by the intense coupling between the bath and the system. 

Figures~\ref{fig:Fig_2}(a-c) demonstrate that MPCs start to emerge when the system-bath coupling strength exceeds a certain threshold. Strong coupling also leads to anomalous effect, resembling overheating and overcooling, driven by energy input from the bath and collective oscillations within the system. As a result, the effective temperature can temporarily drop (rise) below (above) the asymptotic temperature determined by the bath. This feature is robust for a different definition of effective temperature during nonequilibrium processes [see SM]. Notably, when the coupling becomes sufficiently strong, the system statistics are beyond any equilibrium characterization, leading to the emergence of negative temperatures according to the best fit to the fluctuation-dissipation theorem (FDT) [see Figure~\ref{fig:Fig_2}(d)]. Negative temperatures are typically associated with the ability of a system to dissipate heat to all equilibrium systems, including those at $T=\infty$ (where all energy levels are equally populated). The appearance of the transient negative effective temperature is a hallmark of the system's strongly nonequilibrium dynamics. 

\begin{figure}
    \centering
    \includegraphics[width=1.0\columnwidth]{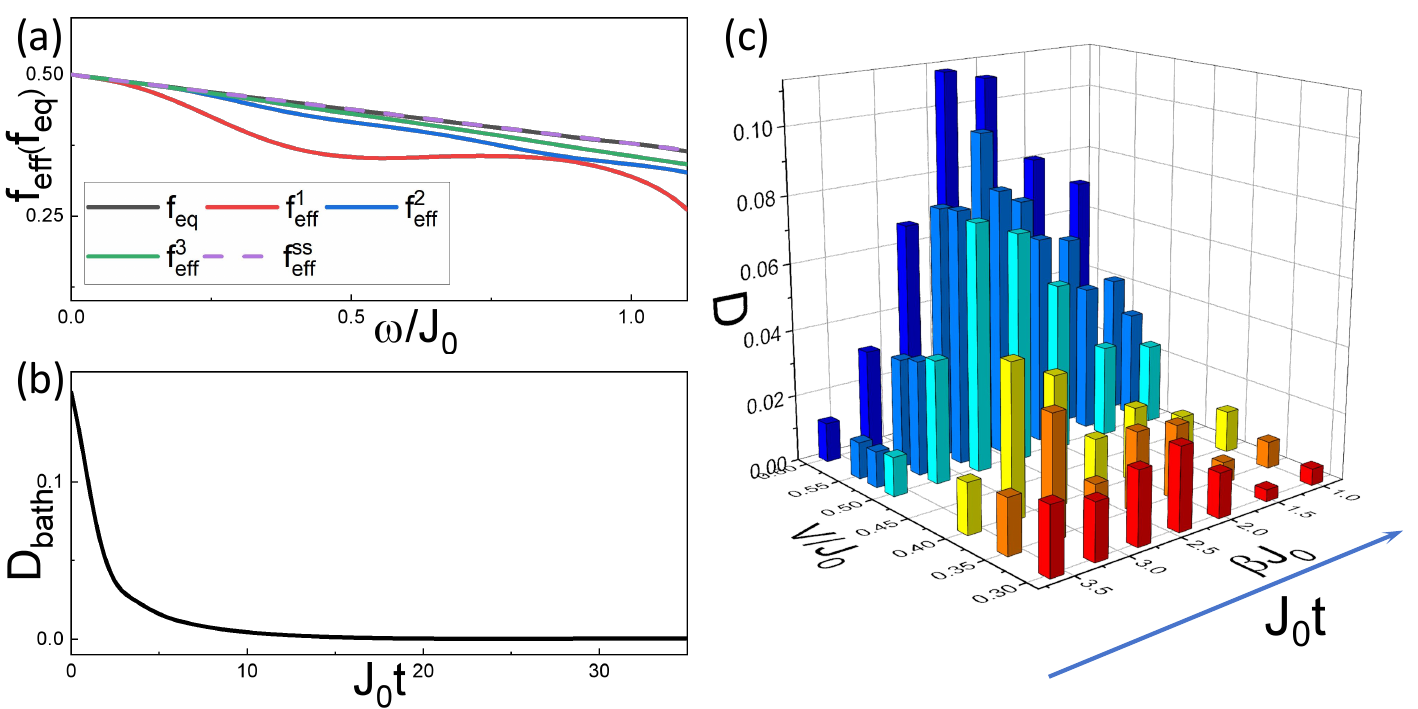}
    \caption{Nonequilibrium statistics during cooling. (a) Energy distribution functions of the system when the effective temperature reaches the bath temperature for the $i$-th time, as indicated in Figure~\ref{fig:Fig_2}(c). (b) Distance of the SYK system during thermalization from  the asymptotic equilibrium state at the bath temperature. The high-frequency cutoff in the computation of the distance function is chosen to be $\omega/J_0=1.0$. (c) Distance function quantifying the deviation from the corresponding time-dependent equilibrium states at the same temperatures during cooling. Parameters are the same as in Figure~\ref{fig:Fig_2}(c), and initial inverse temperature is $\beta_{\rm initial}J_0=4.0$. $J_0$ is the reference coupling strength and is set to be twice the SYK coupling, namely, $J_0=2J$.}
    \label{fig:distribution0_5}
\end{figure}

Figure~\ref{fig:distribution0_5}(a) shows the energy distributions of the fermion modes $f_{\rm eff}^i$ when the system temperatures cross the asymptotic temperature $T_{\rm bath}$ for the $i$-th time, as indicated by the dashed line in Figure~\ref{fig:Fig_2} (c). Here, $f_{\rm eff}^{\rm ss}$ (the dashed purple curve) represents the steady-state distribution at $J_0t=40$ which coincides with the equilibrium distribution given by $f_{\rm eq}(\omega)=\frac{1}{1+e^{\beta_{\rm bath}\omega}}$ (the black curve). While the low-energy behavior of the distribution function $f_{\rm eff}^1$ matches the equilibrium distribution, it deviates at higher energies. As the effective temperature intersects with the bath temperature multiple times, the distribution function gradually approaches the equilibrium form. The nonequilibrium distributions of energy, which change with time, is a demonstration of the nontrivial nonequilibrium dynamics in the SYK model, where the system statistics show complicated behavior beyond equilibrium or analytical formula. The oscillation of effective temperature in Figure~\ref{fig:Fig_2} is the result of the nonequilibrium statistics and definitions of effective temperature beyond equilibrium regimes.

Figure~\ref{fig:distribution0_5}(b) illustrates the time evolution of the distance function, which quantifies the deviation of the energy distribution function from its final-state Fermi-Dirac distribution $f_{\rm bath}(\omega)$ during thermalization, defined by $D_{\rm bath}(t)=\int d\omega |f_{\rm eff}(\omega,t)-f_{\rm bath}(\omega)|$. During the oscillatory period of the system roughly before $J_0 t\lesssim 10$, the distribution function deviates significantly from the steady state and approaches it monotonically over time. 

In contrast, Figure~\ref{fig:distribution0_5}(c) demonstrates the distance between the nonequilibrium distribution function at $\beta_{\rm eff}(t)$ and its equilibrium counterpart at the identical temperatures. The distance measures the degree the system statistics deviate from the equilibrium distribution at the same effective temperature. It is defined as $D(t)=\int d\omega |f_{\rm eff}(\omega,\beta_{\rm eff}(t))-f_{\rm eq}(\omega,\beta_{\rm eff}(t))|$, where $f_{\rm eq}(\omega,\beta_{\rm eff}(t))$ is the Fermi-Dirac distribution at the corresponding effective temperatures. Notably, the deviation of distribution from its equilibrium counterpart is non-monotonic with respect to time during the cooling. Besides, as the system-bath coupling increases, the SYK system is driven away from equilibrium. For SYKs weakly coupled to the bath, the energy distribution of the system during cooling are much closer to the corresponding equilibrium statistics.


\subsubsection{MPCs in SYK models coupled to two baths}
\begin{figure}
    \centering
    \includegraphics[width=1.0\columnwidth]{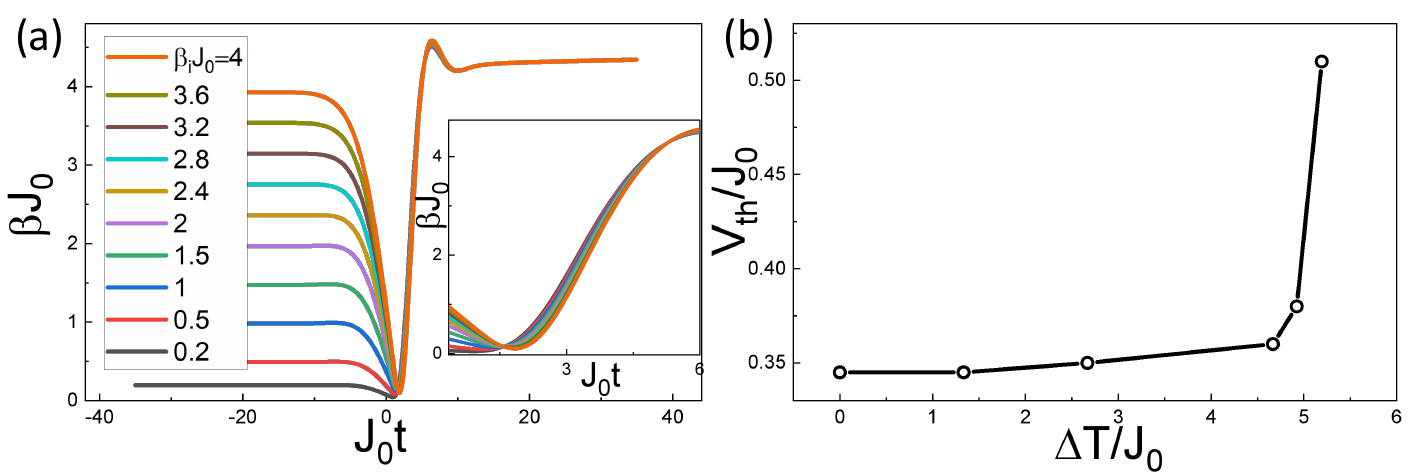}
    \caption{Mpemba crossings in the case of two different baths. (a) The full dynamics from equilibrium states to the steady states. The inset is a zoomed-in view of the dynamics. The parameters used in the numerical simulations are $V_1=V_2=0.5J_0,\, \beta_{bath1}J_0=4.4,\, \beta_{bath2}J_0=4.8$. (b) The threshold coupling $V_\mathrm{th}$ for the emergence of MPCs against the temperature bias between the two baths. For all, $n=3,\, J=0.5J_0.$}
    \label{fig:Fig_4}
\end{figure}

MPCs measured by quantum entanglement and correlations have been shown to be more ubiquitous when coupled to two distinct baths in quantum spin systems \cite{wang2024mpemba}. In weakly interacting $N$-level systems, off-diagonal coherence supported by bath temperature biases can be identified in the quantum master equation, and the enhancement in MPCs has been attributed to initial bath temperature biases and strengthened quantum coherence
\cite{wang2024mpemba,wang2019nonequilibrium,wang2022effect,zhang2020influence}. Whether this holds for strongly coupled SYK systems in the thermalization process measured by effective temperature remains unknown. Employing similar procedure outlined earlier, we numerically solve the dynamics of an SYK model coupled to two distinct baths. We fixed the mean temperature of the two thermal baths, and vary the temperature bias between them. Unlike the enhancement of MPEs in integrable systems such as quantum dots and double-fermionic systems by the bath temperature gradient, bath temperature biases inhibit the MPCs in the SYK model. As demonstrated in Figure~\ref{fig:Fig_4}, stronger bath-system couplings are required for the MPCs to emerge as larger temperature biases are present between baths. This difference serves as one evidence that the underlying mechanism of MPCs in quantum integrable models differs from that in SYK models. In the $N$-level quantum spin systems, the MPC emergence can be determined by the initial settings of the baths, such as their temperatures and temperature biases, as well as the systems' initial spin configurations. However, in the SYK model, the MPC emergence cannot be deduced from those initial conditions, but rather from the coupling strength between the system and the bath. The system-bath coupling strength is the determinant for the rate of cooling or heating of the system.  
From the comparison, we can conclude that the MPCs are dynamical anomalies driven by the nonequilibrium statistics during rapid thermalization in the SYK models. Enlarging the temperature biases leads to complicated averaging effect from the two baths.

\subsubsection{Thermalization of Lindbladian SYK}
\begin{figure}
    \centering
    \includegraphics[width=1.0\columnwidth]{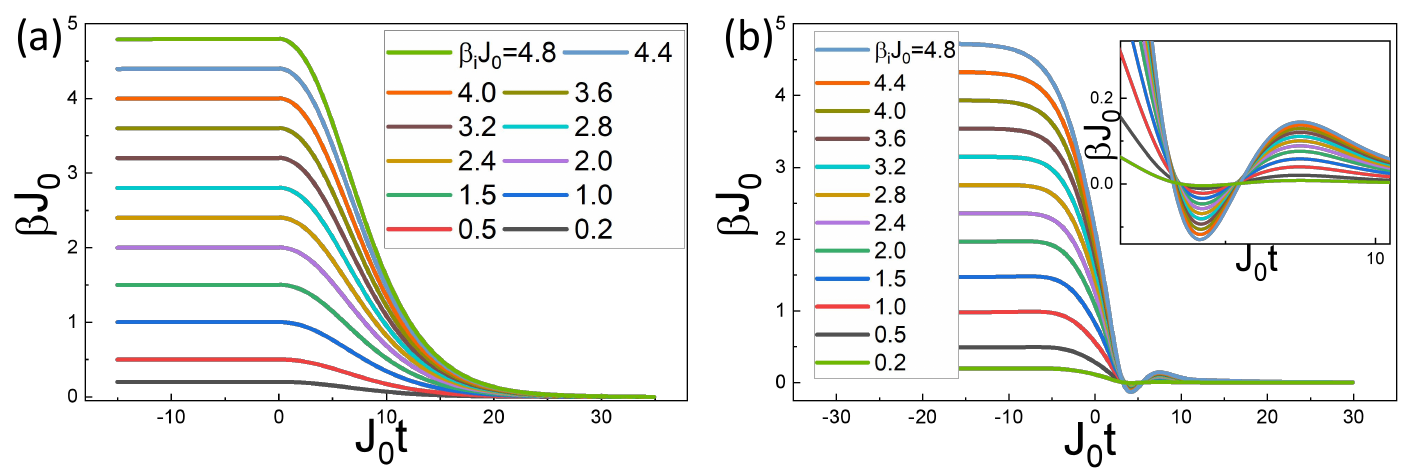}
    \caption{(a) The effective temperature dynamics in the Lindblad description at the dissipative constant $\mu=0.1$. The SYK coupling is set to be $J/J_0=0.5$ and the interacting order is $q=4$. (b) The effective temperature dynamics in the exact calculation of SYKs coupled with baths at infinite temperature. The inset is a zoomed-in view of the dynamics. The parameters used are $J/J_0=0.5,\, V/J_0=0.525, \, n=3$.}
    \label{fig:Fig_5}
\end{figure}

In the studies of open quantum systems, the frequently used method to simulate the dynamics is the Lindblad equation. The Lindbladian formalism provides a simple and tractable description of the bath effect and stands as a good approximation in the weak system-bath coupling regime \cite{breuer2002theory}. For the dissipative SYK model, the Lindbladian given by 
\begin{align}
  \mathcal{L}
  (\rho) = - i [H_{{\rm SYK}}, \rho]  +  \sum_{\alpha} \left[L^{\alpha}\rho L^{\alpha\dag}- \frac{1}{2} \{ L^{\alpha\dag} L^{\alpha}, \rho \} \right]
\end{align}
where $L^\alpha=\sqrt{\mu} \psi^\alpha$ is the linear jump operator. The Lindbladian can be mapped from the density matrix representation to the double-vector representation through Choi-Jamiolkwski isomorphism \cite{sa2022lindbladian,kulkarni2022lindbladian}:
\begin{align}
  \mathcal{L}= -i H_L^{\mathrm{SYK}}-i (-i)^q H_R^{\mathrm{SYK}}-i\mu \sum_i \psi_L^i \psi_R^i -\frac{\mu N}{2}\,.
\end{align}
While exact calculations of the system dynamics coupled to baths reveal the emergence of MPCs, the phenomenological description of SYK dynamics via Lindblad formalism does not manifest similar effect [see Figure~\ref{fig:Fig_5}]. The Lindbladian SYK is an analogy of an SYK model weakly coupled to a bath at infinite temperature. However, exact calculations of an SYK coupled to an SYK bath at $T/J_0=\infty$ show notable differences from the Lindbladian results even when the Lindbladian coupling is strong. This discrepancy can be understood by noticing that the Green's function of the SYK model still has a finite width proportional to the SYK coupling at infinite temperature \cite{zhang2021obstacle}, which is drastically different from the Dirac delta approximation in the Lindbladian method (see Sec.IV in SM \cite{suppl}). Assuming the form of Green's function $G^R(\omega)\approx \frac{1}{\om+i\Gamma}$, the quasi-particle decay rate satisfies $\Gamma \approx \frac{1}{\sqrt{q-1}2^{q/2-2}}J$ at high temperatures. This means that SYK models have nontrivial dynamics even at $T/J=\infty$, in contrast to the flat frequency spectrum in the SYK Lindbadian. The above results demonstrate that the properties as well as different characterizations of the baths can significantly alter the nonequilibrium dynamics after the quench. The failure of Lindbladian dynamics to capture thermalization anomalies suggests that MPCs are potentially more universal in exact calculations than previously recognized using the Lindbladian method.

\subsection{Implications on black hole thermalization}
Modern views based on AdS/CFT correspondence regard black holes as ordinary quantum systems with $A/4G_N$ degrees of freedom--where $A$ is the horizon area--obeying laws of quantum mechanics and thermodynamics \cite{almheiri2021entropy}. Then, the natural question whether phenomena that widely exist across quantum systems also occur in black holes has become a compelling avenue of inquiry. Although the SYK model is known for its ability to bridge the gap between quantum systems and black holes, the duality between SYK and JT gravity is only exact for isolated large-$N$ low-energy dynamics. 
Therefore, the appearance of MPCs in nonequilibrium SYK models serves as an indicator for thermal anomalies in black holes during strongly nonequilibrium processes. If such effect can emerge in black hole thermalization processes, it suggests possible deviations from Hawking's radiation formula during rapid cooling or heating, much like how the Mpemba effect challenges the Stefan–Boltzmann law in its original context, as a purely thermal spectrum does not permit temperature crossings. One concern is that in the expanding universe we live, black holes do not gradually approach the thermal equilibrium with their surrounding baths (e.g., the cosmic background radiation), instead, evolve away from the equilibrium. However, in the early epoch when the universe was hot and opaque, the mean free path of radiated photons were extremely short and black holes with their immediate surroundings formed closed thermal boxes of finite sizes which were approximately non-expanding due to strong gravity and segregated from the far outside regions. 
For instance, a Schwarzchild black hole and its surrounding radiation acquire a stable thermal equilibrium solution under small perturbations in a confined region when the black hole mass $M_{\mathrm{BH}}$ and the radiation energy $U_{\mathrm{R}}$ satisfy $M_{\mathrm{BH}}>4U_{\mathrm{R}}$. 
Nonetheless, critical distinctions between the dissipative SYK models and rapidly-evolving black holes during inflation or the radiation-dominated era must be noted, and direct observation or simulation of the black hole nonequilibrium dynamics is required to confirm these implications. 


\section{Discussion and Conclusion}
During the thermalization processes, the temperature evolution should be interpreted as a measurement of the effective temperature, which reflects certain statistical properties rather than providing a complete description of the system. This interpretation, while seemingly obvious, is essential for understanding the emergence of thermalization anomalies in both SYK systems and the original Mpemba experiment. In a classical Mpemba experiment, the temperature trajectory of the hotter liquid can intersect with that of the colder liquid, but their internal states and statistics may not be identical (as exemplified in Fig.~\ref{fig:distribution0_5}). In fact, for dissipative classical systems, it is known that their internal states can differ even when thermometers register identical temperatures, as exemplified by internal macroscopic flows and heterogeneity in glassy and granular systems \cite{j2007statistical,ritort2003glassy,ediger2000spatially,yamamoto1998dynamics}. Therefore, the system dynamics are not solely determined by the effective temperature during rapid thermalization, and additional parameters are necessary to fully characterize the state of the system. 

In contrast to the quantum MPEs observed in spin models, where crossings are determined primarily by the initial states’ overlaps with slow relaxation modes and the rate of system dynamics plays a minor role, the thermalization MPCs in SYK models are dynamically driven and do not require finely tuned initial conditions. The rapid cooling or heating drives the energy distribution away from the equilibrium Fermi–Dirac form and leads to nontrivial thermalization dynamics of the system. 
The mechanism behind the thermalization MPCs in the SYK models also sheds light to classical MPEs: systems starting from different initial temperatures do not repeat the same cooling paths, which deviate from that predicted by quasi-static thermodynamics. 


Importantly, the emergence of MPCs in the dissipative SYK model is not peculiar to this model, but stems from generic features of strongly coupled, fast-scrambling quantum chaotic systems. The essential ingredient for the MPCs in the SYK model is the dynamical imbalance between the rapid scrambling of correlations and the much slower relaxation of energy modes during nonequilibrium evolution. This pronounced separation of timescales is a hallmark of systems that saturate or approach the chaos bound, \( \lambda_L \lesssim 2\pi k_B T/\hbar \). The scrambling time \( \tau_S \) quantifies how quickly initially localized quantum information spreads across the system and can be extracted from the exponential growth of the out-of-time-order correlator (OTOC) \( \langle [W(t),V(0)]^2 \rangle \sim e^{\lambda_L t} \), where \( \lambda_L \) is the Lyapunov exponent \cite{maldacena2016bound,polchinski2016spectrum,shenker2014multiple}. In contrast, the energy relaxation time \( \tau_E \) characterizes the timescale over which the system exchanges energy with the bath and approaches thermal equilibrium. The strong separation of timescales  \( \tau_S \ll \tau_E \) enables transient population inversions and nonmonotonic temperature trajectories to develop during rapid thermalization. Since this mechanism originates from chaotic information scrambling rather than from model-specific randomness, we expect that it might be universal among strongly interacting, nonintegrable systems that exhibit quantum chaos. By contrast, integrable systems—constrained by an extensive set of conserved quantities and coherent relaxation dynamics—generally lack such a hierarchy of timescales.



In conclusion, nonequilibrium dynamics have manifested many intriguing behaviors beyond conventional thermodynamic predictions in both classical and quantum realms. In particular, the emergence of MPCs in the SYK models showcases the effects of nonequilibrium statistics during thermalization, which are absent in quasi-static systems where thermalization follows a unique trajectory of equilibrium states. This also contrasts with the  MPCs in integrable systems where their emergence depends on the special properties of certain initial states and the bath nonequilibrium conditions. The findings hold potential for experimental verification, with ongoing efforts in simulating SYK models using cold atoms and quantum computers \cite{pikulin2017black,danshita2017creating,luo2019quantum}. For the future outlook, while tracking system temperatures resembles the classical Mpemba experiment, quantities like entanglement entropy and Loschmidt amplitude are also viable parameters to search for potential dynamical anomalies in thermalization \cite{kawabata2023dynamical,chatterjee2023quantum,wang2024mpemba}.  

In addition, thermalization dynamics in SYK models bear significant implications for black hole thermodynamics. Although black holes in our physical universe differ from their AdS counterparts, the observation of MPCs in the SYK model highlights a mechanism of dynamically driven nonequilibrium behavior that may extend far beyond condensed-matter realizations. This connection suggests that the interplay between fast scrambling and slow energy relaxation could also provide insight into nonequilibrium processes in gravitational systems and holographic duals.
Future investigations, through direct numerical simulations of black hole dynamics and observations of temperature oscillations near primordial black holes, are critical to provide more insight than the indications presented in this manuscript. 

\section{Method}

For an interacting SYK model with Hamiltonian given by Eq.~\eqref{eq:H}, the equilibrium correlation function can be solved from the Schwinger-Dyson (SD) equation:
\begin{align}
&G^{-1}_\chi(\omega_n)=-i\omega_n-\Sigma_\chi(\omega_n),\\ &\Sigma_\chi(\tau)=J^2G_\chi(\tau)^3+V^2G_\psi(\tau)^n.
\label{eq:sde}
\end{align}
The diagrammatic representation of the self energy is demonstrated in Figure~\ref{fig:2}(a), where only contributions of order $N^0$ are presented. The solutions to the SD equation can be obtained by iterating Eq.~\eqref{eq:sde} from an non-interacting equilibrium correlator 
$ G_\psi(t,\beta)=b_\psi\frac{\mathrm{sgn}(t)e^{-i\pi/4}\,\pi^{1/2}}{\left(\beta\, \mathrm{sinh}\frac{\pi t}{\beta}\right)^{1/2}}$ \cite{maldacena2016remarks},
and are used as the initial conditions of the SYK model. 

The quench dynamics in real time can be analysed on the Keldysh contour where fields with subscripts ``$-$" signs (e.g. $\psi_-$) live on the lower contour $\mathcal{C}_-$ and fields with subscripts ``$+$" signs live on the upper contour $\mathcal{C}_+$ as shown in Figure~\ref{fig:2}(b). The self-energy can be written as \cite{zhang2019evaporation}
\begin{align}
&\hat\Sigma_{\chi,\alpha\beta}(t,t')\equiv\begin{pmatrix}
\Sigma_\chi^{T}(t,t')&-\Sigma_\chi^{<}(t,t')\\
-\Sigma_\chi^{>}(t,t')&\Sigma_\chi^{\tilde{T}}(t,t')
\end{pmatrix}_{\alpha\beta}\notag\\
&=-J^2\alpha \beta G^3_{\chi,\alpha\beta}(t,t')-V^2\alpha \beta (-1)^{\frac{n+1}{2}}\theta(t)\theta(t')G^{n}_{\psi,\alpha\beta}(t,t')
\end{align}
where $\alpha$ and $\beta$ are $+$ and $-$ signs. The influence of the SYK model on the self energy of the bath is of higher orders in the large-$N$ expansion and can be ignored, leading to the self-energy for the bath unchanged:
\begin{align}
\hat\Sigma_{\psi,\alpha\beta}(t,t')&\equiv\begin{pmatrix}
\Sigma_\psi^{T}(t,t')&-\Sigma_\psi^{<}(t,t')\\
-\Sigma_\psi^{>}(t,t')&\Sigma_\psi^{\tilde{T}}(t,t')
\end{pmatrix}_{\alpha\beta}\notag \\ &=-J^2\alpha \beta G^3_{\psi,\alpha\beta}(t,t').
\end{align}
The correlators defined on the Keldysh contours can be viewed as a two-by-two matrix in the 2D space of the contour branch indices  \cite{babadi2015far} such as
\begin{align}
\hat{G}_{\chi,\alpha\beta}(t,t')=-i\left<\mathbf{T}_c \chi_\alpha(t)\chi_\beta(t')\right>=\begin{pmatrix}
G^{T}_\chi(t,t')&&G^{<}_\chi(t,t')\vspace{3mm}\\ 
G^{>}_\chi(t,t')&&G^{\tilde{T}}_\chi(t,t')
\end{pmatrix}\,,
\label{eq:grn}
\end{align}
where $\alpha,\beta=\pm$, $G^{T}_\chi(t,t')=\theta(t-t')G^{>}_\chi(t,t')+\theta(t'-t)G^{<}_\chi(t,t')$ and $G^{\tilde{T}}_\chi(t,t')=\theta(t'-t)G^{>}_\chi(t,t')+\theta(t-t')G^{<}_\chi(t,t')$.
The symbol $\mathbf{T}_c$ denotes the contour ordering of the  operators such that they are arranged from right to left in the same order as their sequence along the contour. For example, $\mathbf{T}_c \chi(t^+_1) \chi(t^-_2)=\xi \chi(t^-_2) \chi(t^+_1)$, where $\xi=1$ for bosonic operators and $\xi=-1$ for fermionic operators. For states in thermal equilibrium or evolved from equilibrium states, the greater and lesser function satisfies $G^>(t,t')=(G^<(t,t'))^*$ and solving for one of them will suffice \cite{babadi2015far,zanoci2022energy}.

\begin{figure}
    \centering
    \includegraphics[width=1\columnwidth]{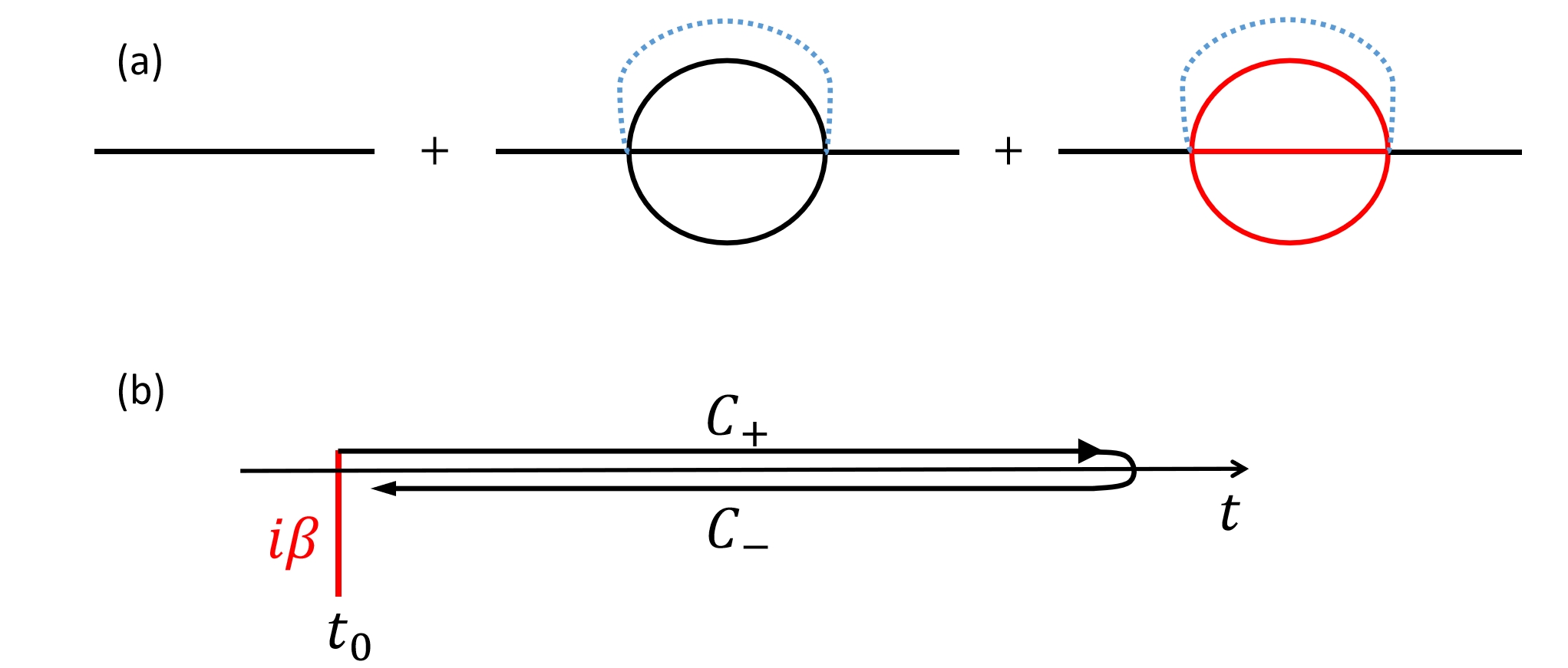}
    \caption{Illustration of the self energy diagram and Keldysh contours. (a) Leading-order corrections to the two-point function of the system SYK$_{\chi}$ for $q=4,\, n=3$. The solid black line represents the correlator of the Majorana fermion $\chi$. The red line represents the correlator of the bath fermion $\psi$. The dotted line represents the disorder averaging that identifies the coupling constants of the connected vertices. The diagrams can be summed by computing the self consistency equations of the propagator and self energy. (b) The upper and lower contours are denoted by $C_+$ and $C_-$, respectively. The thermmal state is prepared at the time $t_0 \ll 0$.}
    \label{fig:2}
\end{figure}

The dynamics of the SYK system after the quench can be described by the Kadanoff-Baym (KB) equations, which we solve numerically. Using the Langreth rule \cite{kamenev2023field}, we integrate the Schwinger-Dyson equation and obtain the Kadanoff-Baym equations for the Green's functions of the SYK system given as follows:
\begin{align}
 i\partial_{t_1}G^>_\chi(t_1,t_2)=\int d t_3 (&\Sigma^R_\chi(t_1,t_3)G^>_\chi(t_3,t_2)\notag \\ &+\Sigma^>_\chi(t_1,t_3)G^A_\chi(t_3,t_2)), \label{eq1}\\
 -i\partial_{t_2}G^>_\chi(t_1,t_2)=\int d t_3 (&G^R_\chi(t_1,t_3)\Sigma^>_\chi(t_3,t_2)\notag \\ &+G^>_\chi(t_1,t_3)\Sigma^A_\chi(t_3,t_2)),\label{eq2}
 \end{align}
where the retarded and advanced self energies (similarly for the Green's functions) are defined as
\begin{align}
\Sigma_{R}(t,t')&=\theta(t-t')(\Sigma^>(t,t')-\Sigma^<(t,t')),\\
\Sigma_{A}(t,t')&=\theta(t'-t)(\Sigma^<(t,t')-\Sigma^>(t,t'))\,.
\end{align}
We numerically solve Eqs.~\eqref{eq1} and \eqref{eq2} by discretizing $(t_1,t_2)$ into a $1000\times 1000$ lattice with size $\Delta t$. For SYKs in contact with two thermal baths and Liouvillian SYKs, similar calculations are performed with the corresponding KB equations. For these two cases, the self energies in the KB equations are modified. The details of these two setups, the method of numerical simulations, as well as the derivations of Lindbladian SYKs are provided in the SM \cite{suppl}.

Given the data of the Green's functions, the temperatures of the system can be obtained. For systems out of thermal equilibrium, temperature is not uniquely defined; however, differences between definitions are usually quantitative and minor. We consider the diagonal slices of the Green's function $G^>(t_1,t_2)$,
\begin{align}
    G_d^>(t;t')=G^>(t+t',t-t')\,,
\end{align}
then apply the fluctuation-dissipation theorem (FDT) \cite{kamenev2023field,kubo1966fluctuation,zhang2021quantum} to find the effective temperature at time $t$:
\begin{align}
    \frac{\mathrm{Im}(G^>(\om,t)+G^<(\om,t))}{\mathrm{Im}(G^R(\om,t))}=-\tanh{\frac{\beta(t) \om}{2}}\,,
\end{align}
where the Green's function $G(\omega,t)$ in the above equation is defined through Wigner transformation 
\begin{gather}
    G(\omega,t)=\int_0^\infty dt' \ e^{i \omega t'} G(t+t'/2,t-t'/2)\,.
    \label{eq:wigner}
\end{gather}
Therefore, the inverse of the effective temperature can be expressed as \cite{eberlein2017quantum}
\begin{gather}
    \beta(t)=\left.\frac{2 \cdot \mathrm{Im}(G_{\chi,K}(\omega,t))}{\omega \cdot \mathrm{Im}(G_{\chi,R}(\omega,t))}\right|_{\omega\rightarrow 0}\,.
    \label{eq:effectivetemp}
\end{gather}
Other ways to quantify the temperature of a system exist, though they are qualitatively similar and converge in thermal equilibrium. These characterizations often relies on the low-frequency limit of the Green's function, and the conclusions of this study are unchanged for such definitions. The results using a different quantification of effective temperature are provided in the SM \cite{suppl}.

\section{Acknowledgement}
X.W. thanks Pengfei Zhang, Tokiro Numasawa and Huajia Wang for valuable discussions, and Pengfei Zhang for assistance with numerical simulations. X.W. also acknowledges the workshop Holography in Beijing 2024 at KITS, UCAS, where part of this work was discussed. X.W. is supported by the FYUST start-up grant, the National Natural Science Foundation of China (NSFC) (Grant No. 12505042), and the NSFC of Zhejiang Province (Grant No. LQN25A050004). J.S. acknowledges support from the NSFC (Grant Nos. 12234019, 12404237) and the NSFC of Zhejiang Province (Grant No. LZ24A040002). The authors acknowledge ChatGPT-5 and DeepSeek-V3 for language polishing of the manuscript. 


\appendix
\onecolumngrid
\clearpage

\section{\Large Supplemental Material: Thermalization and Mpemba-like patterns in effective temperature dynamics of strongly coupled dissipative quantum chaotic systems}

\maketitle

%
%

\bigskip
In the Supplemental Material (SM), we detail the calculations of the Kadanoff-Baym equations for three models: an SYK model coupled to an SYK bath, an SYK model coupled to two distinct SYK baths at different temperatures, and a Lindbladian SYK model. We also provide deeper discussions of the results, including the definitions of effective temperatures in SYK models, total energy dynamics after the quench, and numerical simulations of equilibrium Green's functions and quench dynamics. Additionally, we show that our conclusions hold under different definitions of effective temperature. As noted in the main text, the Lindbladian dynamics show no anomalous behavior, unlike the SYK-bath coupling, and we propose a possible explanation for this discrepancy.


\section{Quantum MPEs and MPCs}
\begin{figure}[b]
    \centering
    \includegraphics[width=0.4\textwidth]{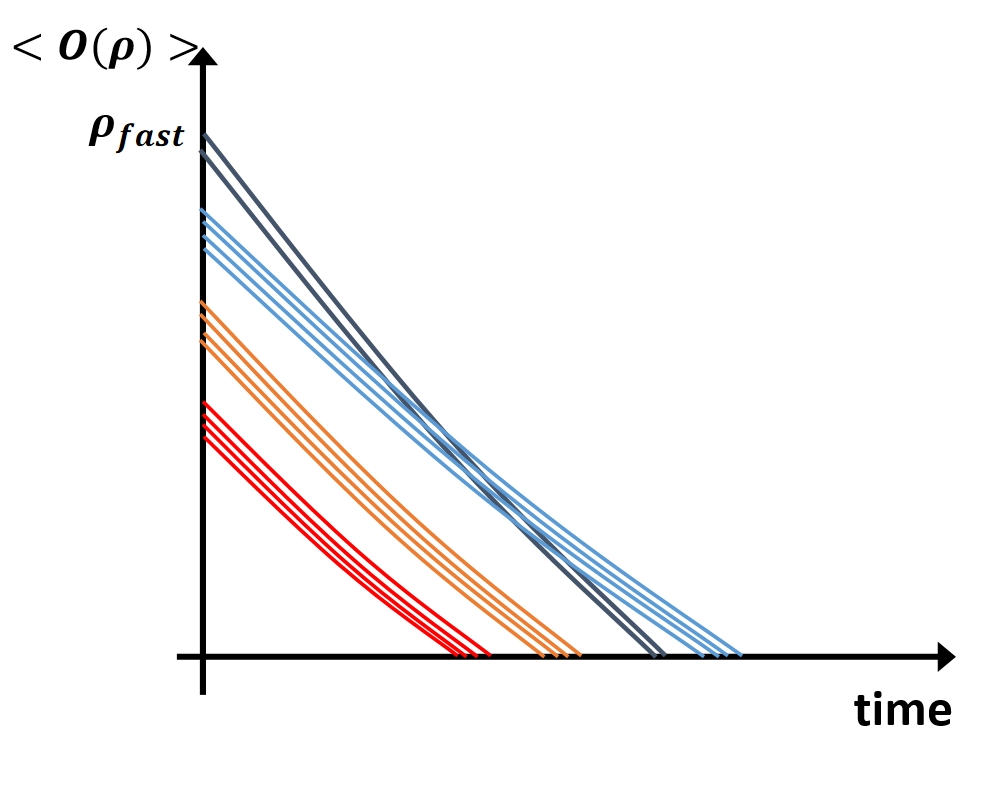}
    \caption{   
    In quantum systems with discrete Liouvillian eigenmodes, the presence of fast-decaying modes (the black curves) determines the emergence of the MPE for the observable $\mathcal{O}(\rho)$. In such cases, the MPE typically depends on whether the initial state largely overlaps with the fast modes. Consequently, the emergence of the MPE relies more on the initial conditions of the state rather than on how rapidly the system is cooled.}
    \label{fig:MPE}
\end{figure}

The anomalous trajectory crossings of chosen measures such as entanglement asymmetry and trace distance  between two initial states have been considered analogous to the classical MPE. Nevertheless, the differences between them should not be overlooked. One can demystify the MPEs in measures such as entanglement symmetry restoration by noticing that the state further from the steady state, as measured by entanglement asymmetry, is not necessarily the state further away from the steady state in the thermalization process [see Fig.~\ref{fig:MPE}]. In fact, the original MPE points to anomalies in thermalization dynamics, while the quantum MPE related to entanglement asymmetry reflects an anomaly in the entanglement measure itself. Specifically, states that thermalize to equilibrium faster can still display larger entanglement asymmetries. This quantum MPE sidesteps the original problem of thermalization dynamics by shifting focus to a different measure--entanglement asymmetry—yet the broader question of anomalous thermalization remains. For the Markovian MPE, it can be attributed to different decay rates of eigen modes and the overlaps between the initial states and the slowest mode. Therefore, in both cases, the emergence of the MPEs relies on the peculiarity of the initial state and the property of the chosen measure. However, unlike in classical systems—where MPEs vanish under infinitely slow cooling and quasi-equilibrium thermodynamics—in quantum systems, the nonequilibrium dynamics are not as decisive. The key difference from the original MPE is that in the classical MPE, if the cooling process is infinitely slow such that the system can be treated by equilibrium thermodynamics, MPEs will not emerge. While in the above cases of the quantum analog, the nonequilibriumness of the system does not play such a decisive role.

Originally, the MPE refers to the emergence of a single anomalous temperature crossing, or an MPC, in the cooling dynamics, so that the relative positions of the effective temperatures are switched when the system is measured at at a designated temperature $T_c$--the icing point in the original Mpemba experiment. Typically, this designated temperature is the temperature of a phase transition or the asymptotic temperature of the steady state. However, for systems driven even further away from the equilibrium than the Mpemba system in its original conception, the temperature trajectories can have additional crossing below $T_c$. Up to now, the concept of the MPE has been generalized to include not only the dynamics that has one or an odd number of MPCs so that the malposition of temperatures can be observed at $T_c$, but also the dynamics that has an even number of MPCs, in which transient anomalies can be detected in the intermediate state rather than at the final state. In fact, MPC is deeply related to the originally proposed MPE in the icing process of hot liquid as in both cases the temperatures in the cooling processes are understood as the effective temperatures and one monitor the effective temperatures during the entire cooling (heating) processes and claim the detection of the MPE once an MPC is found. The MPC is one of the distinctive manifestations of dynamical anomalies beyond the quasi-equilibrium characterization. It is a feature of a strongly nonequilibrium system and is driven by rapid relaxation. The MPC monitors the entire dynamics of the system and provides a refined characterization of the dynamical anomaly beyond the original MPE, which only needs data regarding the initial and the final states. In this study, we argue that the MPCs found in the quantum chaotic systems share more similarities with the original MPE and originate from the same underlying mechanism in contrast to quantum MPEs found in many quantum integrable models. 

\section{SYK models in thermal baths}
\subsection{SYK model in a thermal bath}
\begin{figure}
    \centering
    \includegraphics[width=0.75\textwidth]{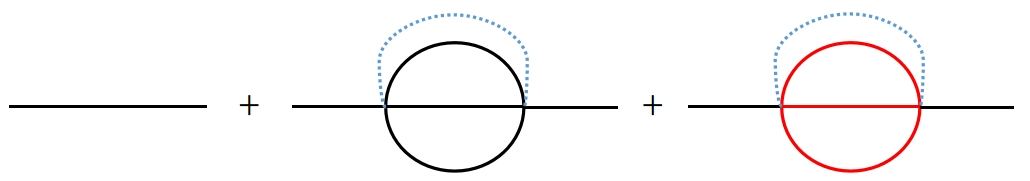}
    \caption{Leading-order corrections to the two-point function of the system SYK$_{\chi}$ for $q=4,\, n=3$. The solid black line represents the correlator of the Majorana fermion $\chi$. The red line represents the correlator of the bath fermion $\psi$. The dotted line represents the disorder averaging that identifies the coupling constants of the connected vertices. The diagrams can be summed by computing the self consistency equations of the propagator and self energy. The self energy $\Sigma$ includes all the one particle irreducible contributions to the propagator.}
    \label{fig:diagram}
\end{figure}

We consider a q-body interacting SYK$_q$ model in a thermal bath, consisting of $N$ Majorana fermions in (0+1) dimensions with random $q$-fermion interactions. The bath is also an SYK$_q$ model governed by the same Hamiltonian with $N^2$ Majorana modes. In the large-$N$ and weak system-bath coupling limits, the model exhibits maximal chaos for $q>2$, saturating the bound on the quantum Lyapunov exponent derived from out-of-time-ordered correlators. For simplicity, we focus on $q=4$ case and the total Hamiltonian is given by
\begin{align}
&H=H_{\text{SYK}}[J_{i_1i_2i_3i_4},\chi]+H_{\text{SYK}}[\tilde{J}_{i_1i_2i_3i_4},\psi]\notag\\&\ \ \ +\sum_{ai_1i_2...i_n}\frac{V_{ai_1i_2...i_n}}{n!}\chi_a\psi_{i_1}\psi_{i_2}...\psi_{i_n}.\label{eq:H}
\end{align}
where $\chi_i$ are the Majorana fermions of the system on site $i=1... N$ obeying 
\begin{align}
\{\chi_i,\chi_j\}=\delta_{ij}\,,    
\end{align}
and $\psi_i$ represent the bath operators on site $i=1... N^2$. $H_{\text{SYK}}[J_{i_1i_2i_3i_4},\chi]$ and $H_{\text{SYK}}[\tilde J_{i_1i_2i_3i_4},\psi]$ are the standard SYK$_4$ Hamiltonians for the system ($\chi$) and the bath ($\psi$), respectively, which are given by
\begin{align}
H_{\text{SYK}}[J_{i_1i_2i_3i_4},\chi]=\sum_{i_1i_2i_3i_4}\frac{J_{i_1i_2i_3i_4}}{4!}\chi_{i_1}\chi_{i_2}\chi_{i_3}\chi_{i_4}\,,\\
H_{\text{SYK}}[\tilde J_{i_1i_2i_3i_4},\psi]=\sum_{i_1i_2i_3i_4}\frac{\tilde J_{i_1i_2i_3i_4}}{4!}\psi_{i_1}\psi_{i_2}\psi_{i_3}\psi_{i_4}\,.
\label{1}
\end{align}
The distributions of the interaction strengths are given by:
\begin{align}
&\overline{J_{i_1i_2i_3i_4}}=0,\ \ \ \ \ \ \overline{\tilde{J}_{i_1i_2i_3i_4}}=0,\ \ \ \ \ \ \overline{V_{ai_1i_2...i_n}}=0,\\
&\overline{J_{i_1i_2i_3i_4}^2}=\frac{3!J^2}{N^3},\ \ \ \overline{\tilde{J}_{i_1i_2i_3i_4}^2}=\frac{3!J^2}{N^6},\ \ \ \overline{V_{ai_1i_2...i_n}^2}=\frac{n!V^2}{N^{2n}}.
\end{align}
For a non-interacting SYK model in equilibrium with the time translation symmetry, the Euclidean Green's function at the zero-temperature limit is
\begin{equation}
G_\psi(\tau)=\langle T (\psi(\tau)\psi(0)) \rangle=\langle \psi(\tau)\psi(0) \rangle \theta(\tau)-\langle \psi(0)\psi(\tau) \rangle \theta(-\tau)=b_\psi\frac{\mathrm{sgn}(\tau)}{|\tau|^{1/2}}, \ \quad 4\pi J^2 b_\psi^4=1,\,
\end{equation}
and the Green's function at finite temperature is given by conformal mapping $\tau=\tan\frac{\pi\tau'}{\beta}$ \cite{maldacena2016remarks}. This gives the correlation function at the inverse temperature $\beta$
\begin{align}
    G_\psi(\tau)=b_\psi\frac{\mathrm{sgn}(\tau)\,\pi^{1/2}}{\left(\beta\, \mathrm{sin}\frac{\pi \tau}{\beta}\right)^{1/2}}, \ \quad 4\pi J^2 b_\psi^4=1\,.
\end{align}
From the Euclidean correlators, we can obtain the zero- and finite-temperature correlators in Lorentzian time by setting $\tau=it$ \cite{maldacena2016remarks}, which gives
\begin{align}
    G_\psi(t)=b_\psi \frac{e^{-i\pi/4}}{t^{1/2}}\mathrm{sgn}(t),\qquad  G_\psi(t,\beta)=b_\psi\frac{\mathrm{sgn}(t)e^{-i\pi/4}\,\pi^{1/2}}{\left(\beta\, \mathrm{sinh}\frac{\pi t}{\beta}\right)^{1/2}}\,,
\end{align}
where $ G_\psi(t)$ is the Lorentzian correlator at zero temperature and $G_\psi(t,\beta)$ is the Lorentzian correlator at temperature $\beta^{-1}$. For the interacting SYK model with Hamiltonian given by Eq.~\eqref{eq:H}, the Green's function can be read directly from Fig.~\ref{fig:diagram}:
\begin{align}
&G^{-1}_\chi(\omega_n)=-i\omega_n-\Sigma_\chi(\omega_n),\\ &\Sigma_\chi(\tau)=J^2G_\chi(\tau)^3+V^2G_\psi(\tau)^n.
\end{align}
In the diagrammatic representation Fig.~\ref{fig:diagram}, only contributions of order $N^0$ are presented. 

\begin{figure}[t]
    \centering
    \includegraphics[width=0.46\textwidth]{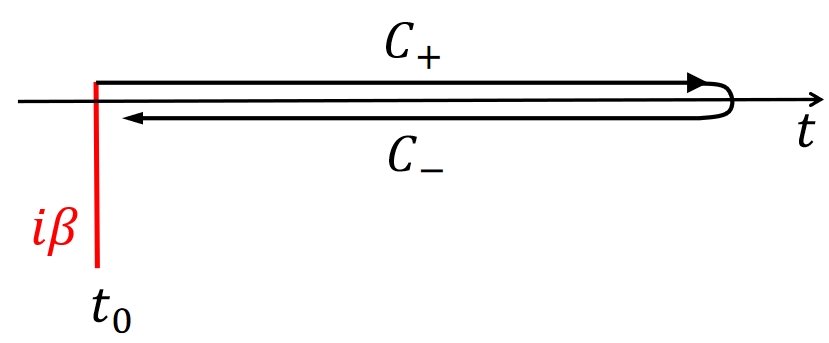}
    \caption{Demonstration of the Keldysh contour used in this study. The upper and lower contours are denoted by $C_+$ and $C_-$, respectively. The thermmal state is prepared at the time $t_0 \ll 0$.}
    \label{fig:kcontour}
\end{figure}

In real time, the quench dynamics can be analysed on the Keldysh contour where fields with subscripts ``$-$" signs (e.g. $\psi_-$) live on the lower contour $\mathcal{C}_-$ and fields with subscripts ``$+$" signs live on the upper contour $\mathcal{C}_+$ as shown in Fig.~\ref{fig:kcontour}. In this case, the self-energy for the system can be written as
\begin{align}
\hat\Sigma_{\chi,\alpha\beta}(t,t')&\equiv\begin{pmatrix}
\Sigma_\chi^{T}(t,t')&-\Sigma_\chi^{<}(t,t')\\
-\Sigma_\chi^{>}(t,t')&\Sigma_\chi^{\tilde{T}}(t,t')
\end{pmatrix}_{\alpha\beta}\notag\\
&=-J^2\alpha \beta G^3_{\chi,\alpha\beta}(t,t')-V^2\alpha \beta (-1)^{\frac{n+1}{2}}\theta(t)\theta(t')G^{n}_{\psi,\alpha\beta}(t,t'),
\end{align}
where $\alpha$ and $\beta$ are $+$ and $-$ signs. Similarly, the self-energy for the bath is:
\begin{align}
\hat\Sigma_{\psi,\alpha\beta}(t,t')&\equiv\begin{pmatrix}
\Sigma_\psi^{T}(t,t')&-\Sigma_\psi^{<}(t,t')\\
-\Sigma_\psi^{>}(t,t')&\Sigma_\psi^{\tilde{T}}(t,t')
\end{pmatrix}_{\alpha\beta}\notag \\ &=-J^2\alpha \beta G^3_{\psi,\alpha\beta}(t,t').
\end{align}

The correlators defined on the Keldysh contours can be viewed as a two-by-two matrix in the 2D space of the contour branch indices  \cite{babadi2015far} such as
\begin{align}
\hat{G}_{\chi,\alpha\beta}(t,t')=-i\left<\mathbf{T}_c \chi_\alpha(t)\chi_\beta(t')\right>=\begin{pmatrix}
G^{T}_\chi(t,t')&&G^{<}_\chi(t,t')\vspace{3mm}\\ 
G^{>}_\chi(t,t')&&G^{\tilde{T}}_\chi(t,t')
\end{pmatrix}\,,
\label{eq:grn}
\end{align}
where $\alpha,\beta=\pm$, $G^{T}_\chi(t,t')=\theta(t-t')G^{>}_\chi(t,t')+\theta(t'-t)G^{<}_\chi(t,t')$ and $G^{\tilde{T}}_\chi(t,t')=\theta(t'-t)G^{>}_\chi(t,t')+\theta(t-t')G^{<}_\chi(t,t')$.
The symbol $\mathbf{T}_c$ denotes the contour ordering of the  operators such that they are arranged from right to left in the same order as their sequence along the contour. For example, $\mathbf{T}_c \chi(t^+_1) \chi(t^-_2)=\xi \chi(t^-_2) \chi(t^+_1)$, where $\xi=1$ for bosonic operators and $\xi=-1$ for fermionic operators. 

From the above definition, it is easy to read the averaged ``greater" and ``lesser" Green's functions, which are the correlators of the fields living on two different contours defined as follows:
\begin{align}
    G^>(t_1,t_2)&\equiv G(t^-_1,t^+_2)=\frac{-i}{N}\sum_{i=1}^N \langle \psi(t_1^-)\psi(t_2^+)\rangle \\
    G^<(t_1,t_2)&\equiv G(t^+_1,t^-_2)=\frac{i}{N}\sum_{i=1}^N \langle\psi(t_2^-) \psi(t_1^+)\rangle\,.
\end{align}
The lesser and greater correlation functions are independent functions for Dirac (complex) fermions. For the Majorana fermions, it is straightforward from above equations that 
\begin{align}
    G^>(t,t')=-G^<(t',t)\,,
\end{align}
which holds even for nonequilibrium dynamics. For states in thermal equilibrium or in general states evolved from thermal equilibrium states, the greater and lesser function satisfies $G^>(t,t')=(G^<(t,t'))^*$ \cite{babadi2015far,zanoci2022energy}. Therefore, $G^>(t_1,t_2)=-\left(G^>(t_2,t_1)\right)^*$. In this formalism, sometimes it is more convenient to write the equations in the Keldysh basis, where the retarded, advanced and Keldysh correlators are defined as follows:
\begin{align}
G_{R}(t,t')&=\theta(t-t')(G^>(t,t')-G^<(t,t')),\\
G_{A}(t,t')&=\theta(t'-t)(G^<(t,t')-G^>(t,t')),\\
G_{K}(t,t')&=G^<(t,t')+G^>(t,t').
\end{align}
The above definitions will be used later.

\subsection{Equilibrium solutions}
In this section, we briefly summarize the procedures to obtain the exact equilibrium Green's function of the system when interactions with the bath considered. The Green's function is obtained by solving the Dyson's equation self-consistently starting from the initial input which is set to the conformal limit of the thermal Green's function. One may refer to ref.~\cite{eberlein2017quantum} for details.  

Since the equilibrium solution of the Green's function has the time translational symmetry, the Green's function $G_{\alpha,\beta}(t_1,t_2)$ is only dependent on the relative time $t=t_1-t_2$ and is abbreviated as $G_{\alpha,\beta}(t)$. 
In thermal equilibrium, the Kubo–Martin–Schwinger condition gives \cite{kamenev2023field}
\begin{align}
    G^>(\omega)=\pm e^{\beta \omega} G^<(\omega)\,,
\end{align}
where the $+$ sign is for bosons and $-$ sign is for fermions. The spectral function 
\begin{align}
    A(\om)=-2\mathrm{Im} G_R(\omega)
    \label{eq:spectr}
\end{align}
can be determined by the self-consistent equation of retarded Green's functions. In thermal equilibrium, the greater Green's function is related to the spectral function by
\begin{align}
    G^>(\omega)=-i (1-n_F(\om,T)) A(\om) =-i n_F(-\omega,T) A(\om)\,,
    \label{eq:glw}
\end{align}
where $n_F(\om,T)$ is the Fermi-Dirac distribution function given by $n_F=\frac{1}{1+e^{\beta \om}}$. For Majorana fermions, the operators satisfy $\psi^\dag=\psi$ and the Green's function satisfies:
\begin{align}
    G(-t)=-G^*(t)\,,
\end{align}
one can obtain the retarded Green's function in the conformal limit \cite{maldacena2016remarks}:
\begin{align}
    i\,G^R(t)=\theta(t) \langle [(\psi(t),\psi(0))] \rangle=2b\cos(\pi/q) \,\left(\frac{\pi}{\beta \sinh(\frac{\pi t}{\beta})}\right)^{\frac{2}{q}} \theta(t)\,,
    \label{eq:grtherm}
\end{align}
where time translational symmetry is assumed. The retarded Green's function in Eq.~\eqref{eq:grtherm} can be used as a proper choice of the initial data for generating the exact Green's function through the iteration to be described below. Adopting the following convention for the Fourier transformations:
\begin{align}
    &A(\om)=\int_{-\infty}^\infty dt A(t) e^{i\om t}\,,\\
    &A(t)=\int_{-\infty}^\infty \frac{d\om}{2\pi} A(\om) e^{-i\om t}\,,
\end{align}
we have the self-consistency equations:
\begin{equation}
\begin{split}
    &G_{R}(\omega)^{-1}=\omega-\Sigma_{R}(\omega), \\
    &\Sigma_{R}(\omega)=-iJ^2\int_0^\infty dt e^{i\omega t}(n(t)^3+(n(t)^*)^3), \\
    &n(t)=\int \frac{d\omega}{2\pi} e^{-i\omega t} A(\omega)n_F(\omega, T)\,.
\end{split}
\label{eq:consistencyeq}
\end{equation}
The exact Green's function of the SYK model in thermal equilibrium can be computed numerically by iterating Eqs.\eqref{eq:spectr}, \eqref{eq:glw} and \eqref{eq:consistencyeq} until the solution converges. The solutions are used for the initial conditions of the SYK system and the baths. 

\subsection{Effective temperature and nonequilibrium statistics}
To capture the statistical property of the out-of-equilibrium system after the quench, we track the effective temperature changes with time. For systems out of thermal equilibrium, temperature is not evidently and uniquely defined as in equilibrium systems. Instead, observables such as the effective temperatures are defined to reflect the overall statistical properties of the system and serve as trackers for the possible emergence of the MPE. We consider the diagonal slices of the Green's function $G^>(t_1,t_2)$,
\begin{align}
    G_d^>(t;t')=G^>(t+t',t-t')\,,
\end{align}
then apply the fluctuation-dissipation theorem (FDT) \cite{kamenev2023field,kubo1966fluctuation,zhang2021quantum} to find the effective temperature at time $t$:
\begin{align}
    \frac{\mathrm{Im}(G^>(\om,t)+G^<(\om,t))}{\mathrm{Im}(G^R(\om,t))}=-\tanh{\frac{\beta(t) \om}{2}}\,,
\end{align}
where the Green's function $G(\omega,t)$ in the above equation is defined through Wigner transformation \footnote{Here, we make a small comment about the different conventions as pointed out by Pengfei Zhang. The effective temperature are sometimes written as \cite{zhang2019evaporation}
\begin{gather}
    \beta(t)=2\frac{d}{d\omega}\left(\frac{G_{\chi,K}(\omega,t)}{G_{\chi,R}(\omega,t)-G_{\chi,A}(\omega,t)}\right)_{\omega=0}\,.
    \label{eq:wrongt}
\end{gather}
In this convention, we remind that the Wigner transformation
\begin{gather}
    G(\omega,t)=\int_{-\infty}^\infty dt' \ e^{i \omega t'} G(t+t'/2,t-t'/2)
\end{gather}
has different integration limits from Eq.~\eqref{eq:wigner}.}
\begin{gather}
    G(\omega,t)=\int_0^\infty dt' \ e^{i \omega t'} G(t+t'/2,t-t'/2)\,.
    \label{eq:wigner}
\end{gather}
Therefore, the inverse of the effective temperature can be expressed as \cite{eberlein2017quantum}
\begin{gather}
    \beta(t)=\left.\frac{2 \cdot \mathrm{Im}(G_{\chi,K}(\omega,t))}{\omega \cdot \mathrm{Im}(G_{\chi,R}(\omega,t))}\right|_{\omega\rightarrow 0}\,.
    \label{eq:effectivetemp}
\end{gather}
This definition preserves the FDT but suffers from the possible ailment of causality violation in time \footnote{This is also pointed out in Ref.~\cite{almheiri2024universal}}. Another possible way to characterize the transient effective temperature of the system after the quench is through the corner slice of the Green's function \cite{almheiri2024universal}:
\begin{align}
    G_c^>(t,t')=\theta(t')G^>(t-t',t)+\theta(-t')G^>(t,t+t')\,.
\end{align}
Then, we can similarly define $G_c^>(\om,t)$ by Fourier transforming $G_c^>(t,t')$ on $t'$, i.e.,
\begin{align}
    G_c^>(\om,t)=\int_{-\infty}^\infty dt' \ e^{i \omega t'} G(t,t')\,.
\end{align}
One can then define the inverse temperature by inserting the above definition of $G_c^{>(<)}(\omega,t)$ into the expression below \footnote{Note that the difference between this equation and Eq.~\eqref{eq:effectivetemp} is due to the different convention used in the fluctuation-dissipation relation.}
\begin{align}
      \beta'(t)=\left.\frac{-\mathrm{Im}(G_{\chi,K}(\omega,t))}{\omega \cdot \mathrm{Im}(G_{\chi,R}(\omega,t))}\right|_{\omega\rightarrow 0}\,. \label{eqbeta2}
\end{align}
As remarked in ref.~\cite{almheiri2024universal}, this definition respects the causality but at the expense of the FDT violation at higher frequencies. For this study, we ignore possible issues with the characterization of the effective temperatures of nonequilibrium systems and treat them as certain statistical measures to detect the emergence of the MPE. We remark that these two quantities may differ slightly when the system is away from the equilibrium but qualitatively very similar and agree exactly when the system is in thermal equilibrium. The conclusions in this study are independent of which definition we use. For simplicity, the definition by Eq.~\eqref{eq:effectivetemp} is what we will mainly refer to in this study. 

The time-dependent energy distribution function is defined by extending the equilibrium distribution, which is given by
\begin{align}
    f(\omega,t)={\mathrm{Re}}\left(\frac{-i G^<(\omega,t)}{A(\omega,t)}\right)\,.
    \label{eq:efff}
\end{align}
In the equilibrium scenario, the distribution function follows the Fermi-Dirac distribution. In a relaxation process, the energy distribution generally deviates from the standard equilibrium distribution depending on the degree of nonequilibrium. In this case, Eq.~\eqref{eq:efff} can be viewed as the effective distribution of energy. 

\subsection{Kadanoff-Baym equation and quench dynamics}
The dynamics of the SYK system after the quench can be described by the Kadanoff-Baym equations, i.e., the equations of motion of the system. In these equations, the influence of the bath on the SYK model is of order one, while the influence of the SYK model on the self energy of the bath is of higher orders in the large-$N$ expansion and can be ignored. We solve the Kadanoff-Baym equation numerically. For the SYK systems in the thermal bath, we need to solve for $G_\chi^{>}(t,t')$ and $G_\chi^{<}(t,t')$. For $t,t'<0$, the system is prepared to a thermal equilibrium state and the Green's functions satisfy, viz., $G_\chi^{>}(t,t')=G_\chi^{>}(t-t')$, where we have assumed the time translational invariance. The data can be easily discretized and stored in terms of $\{{\Delta t}, G_\chi^{>}(\Delta t)$\} as the initial condition for numerical simulations of the quench dynamics. 

The retarded, advanced and Keldysh components of the self-energy are defined in a similar way as that for the Green's function, viz.,
\begin{align}
\Sigma_{R}(t,t')&=\theta(t-t')(\Sigma^>(t,t')-\Sigma^<(t,t')),\\
\Sigma_{A}(t,t')&=\theta(t'-t)(\Sigma^<(t,t')-\Sigma^>(t,t')),\\
\Sigma_{K}(t,t')&=\Sigma^<(t,t')+\Sigma^>(t,t').
\end{align}
For the Majorana fermions, the greater and lesser Green's functions satisfy the relation $G^>(t,t')=(G^<(t,t'))^*$. Using the Langreth rule \cite{kamenev2023field}, we integrate the Schwinger-Dyson equation and obtain the Kadanoff-Baym equations for the Green's functions of the SYK system given as follows:
\begin{align}
 i\partial_{t_1}G^>_\chi(t_1,t_2)=\int d t_3 (&\Sigma^R_\chi(t_1,t_3)G^>_\chi(t_3,t_2)\notag \\ &+\Sigma^>_\chi(t_1,t_3)G^A_\chi(t_3,t_2)), \label{eq1}\\
 -i\partial_{t_2}G^>_\chi(t_1,t_2)=\int d t_3 (&G^R_\chi(t_1,t_3)\Sigma^>_\chi(t_3,t_2)\notag \\ &+G^>_\chi(t_1,t_3)\Sigma^A_\chi(t_3,t_2)),\label{eq2}
 \end{align}
where the self-energies for the SYK$_4$ system denoted by $\chi$ and for the bath denoted by $\psi$ are defined as follow:
\begin{align}
\Sigma_\chi^{>}(t,t') &= -J^2 (G^>_{\chi}(t,t'))^3-V^2(-1)^{\frac{n+1}{2}}\theta(t)\theta(t')(G^>_{\psi}(t,t'))^n\,,\\
\Sigma_\chi^{<}(t,t') &= -J^2 (G^<_{\chi}(t,t'))^3-V^2(-1)^{\frac{n+1}{2}}\theta(t)\theta(t')(G^<_{\psi}(t,t'))^n\,,\\
\Sigma_\psi^{<}(t,t')&=-J^2 (G^<_{\psi}(t,t'))^3\,,\\
\Sigma_\psi^{>}(t,t')&=-J^2 (G^>_{\psi}(t,t'))^3\,.
\label{Eq:sigmas}
\end{align}

Then, we numerically solve Eqs.~\eqref{eq1} and \eqref{eq2} by discretizing $(t_1,t_2)$ into a $1000\times 1000$ lattice with size $\Delta t$. The differential equations become the difference equations and the integration can be approximated by summation, and $G^>_\chi$ in the steady state can be simulated by the iterative method. First, we need to get the initial state at each different initial temperature $1/\beta_i$, where the data of on-lattice $G^>_\chi(t_1,t_2)$ can be obtained from the equilibrium solution. Then the difference of $G^>_\chi$ on each lattice ($\Delta_{t_1}G^>_\chi$, $\Delta_{t_2}G^>_\chi$) can be calculated from the summation approximated by Eqs.~\eqref{eq1} and \eqref{eq2}, and we can naturally update the data of $G^>_\chi$ based on the difference. Repeat calculations of the difference and updating of $G^>_\chi$ until $G^>_\chi$ maintains unchanged with time, i.e., the difference of $G^>_\chi$ decay to $0$. In simulation, we choose $J=0.5$ and the cutoff in the time domain $\Lambda_t=50$, and thus restrict $t$ to $[-\Lambda_t,\Lambda_t]$ so that $\Delta t=0.1$. To ensure the accuracy of the simulated results, especially to avoid the size effects, we have augmented the number of lattice to $2000\times2000$. Furthermore, we have modulated the parameter $\Lambda_t$ to either $100$ (corresponding to $\Delta t = 0.1$) or maintained it at $50$ (corresponding to $\Delta t = 0.05$). The quantitative results obtained under these parameter settings exhibit a high degree of consistency with those derived using the selected parameters, thereby validating the accuracy of our simulations. Then, we use the data of the Green's function to compute the dynamics of the inverse of the effective temperature.

\subsection{Numerical results for SYKs in a thermal bath}
It is usually expected that during cooling processes, the temperature of a SYK model exponentially approaches its asymptotic temperature which is determined by its environment. However, we find that the system temperature after the quench shows an oscillatory feature on top of the exponential decay to its asymptotic temperature. This can cause the temperature of two initial states to cross at finite time at strong system-bath couplings. This phenomenon is referred to as the Mpemba crossing. The Mpemba crossing describes a phenomenon that for different initial states the system observables, which are conventionally considered as state functions in a equilibrium setup, intersect at a finite time. The consequence of this crossing is that for an experimenter Alice who prepares two different systems immersed in the same thermal bath and constantly monitors an observable, for example, the effective temperature, Alice may find that the system with a higher initial temperature cools to a temperature lower than that of the system with a lower initial temperature at certain times during the continuous monitoring of the system temperatures. This is a reminiscence of the original Mpemba effect and falls into the current category of the general MPEs.

\begin{figure}
    \centering
    \includegraphics[width=0.43\textwidth]{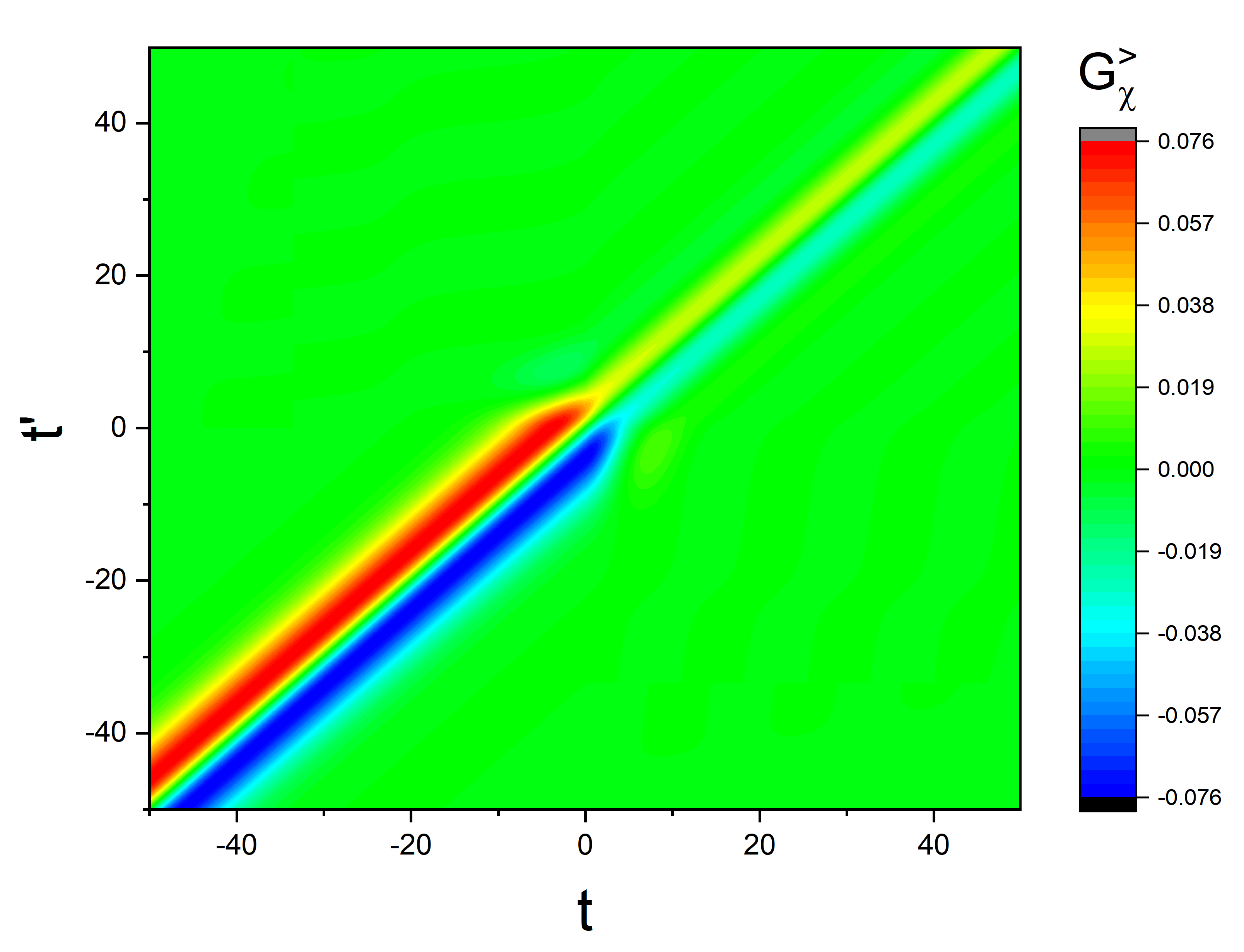}
    \includegraphics[width=0.43\textwidth]{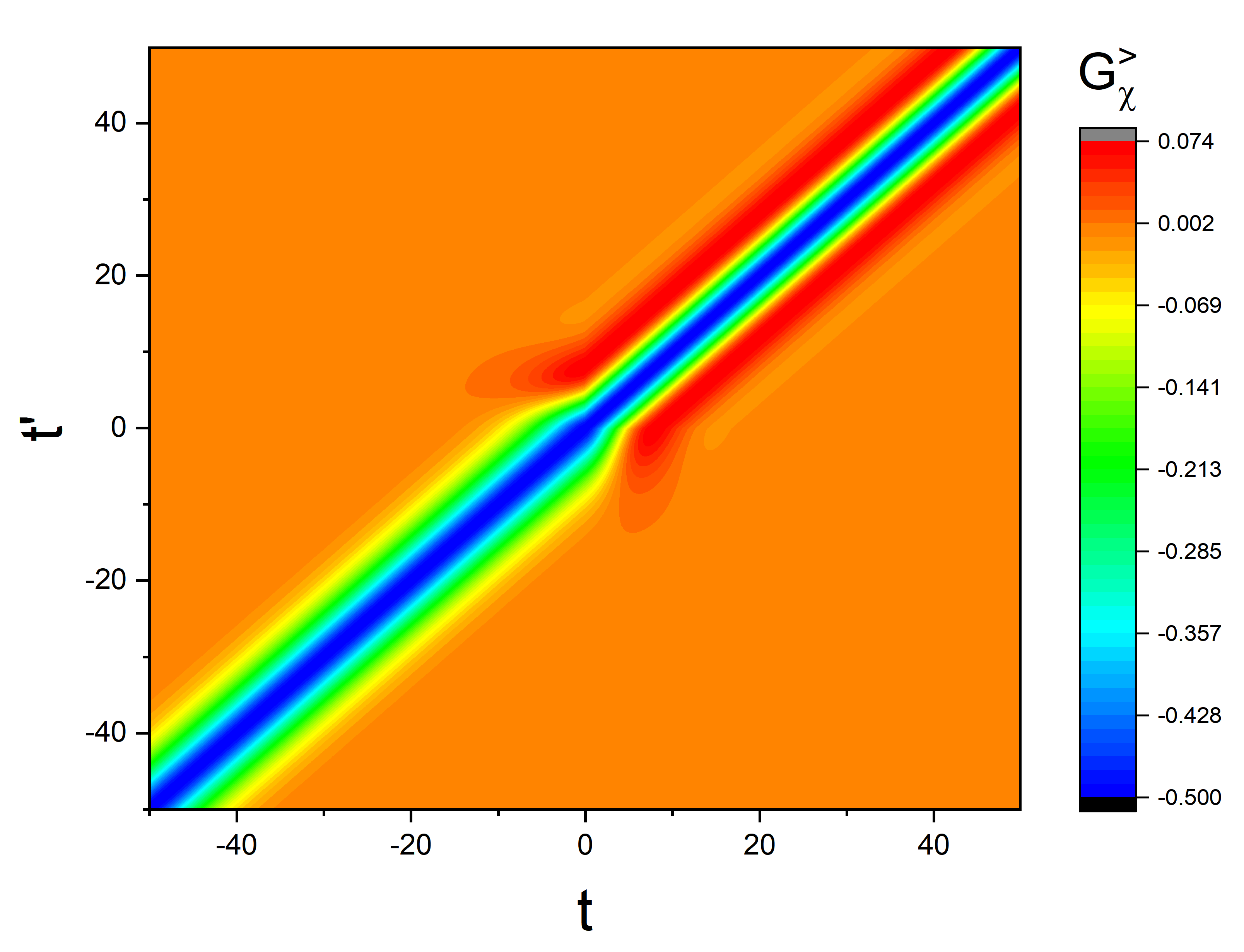}
    \caption{The 2D map for the Green's function $G^>_{\chi}$. (a) The real component. (b) The imaginary component. The coupling between the system and the bath is set to be $V=0.525.$ The other parameters used in the numerical simulations are: $n=3,\, J=0.5,\, \beta_{\mathrm{bath}}=0.5,\, \beta_{i}=2.4,\, \Delta t=0.1$.}
    \label{fig:grn1}
\end{figure}

\begin{figure}
    \centering
    \includegraphics[width=0.45\textwidth]{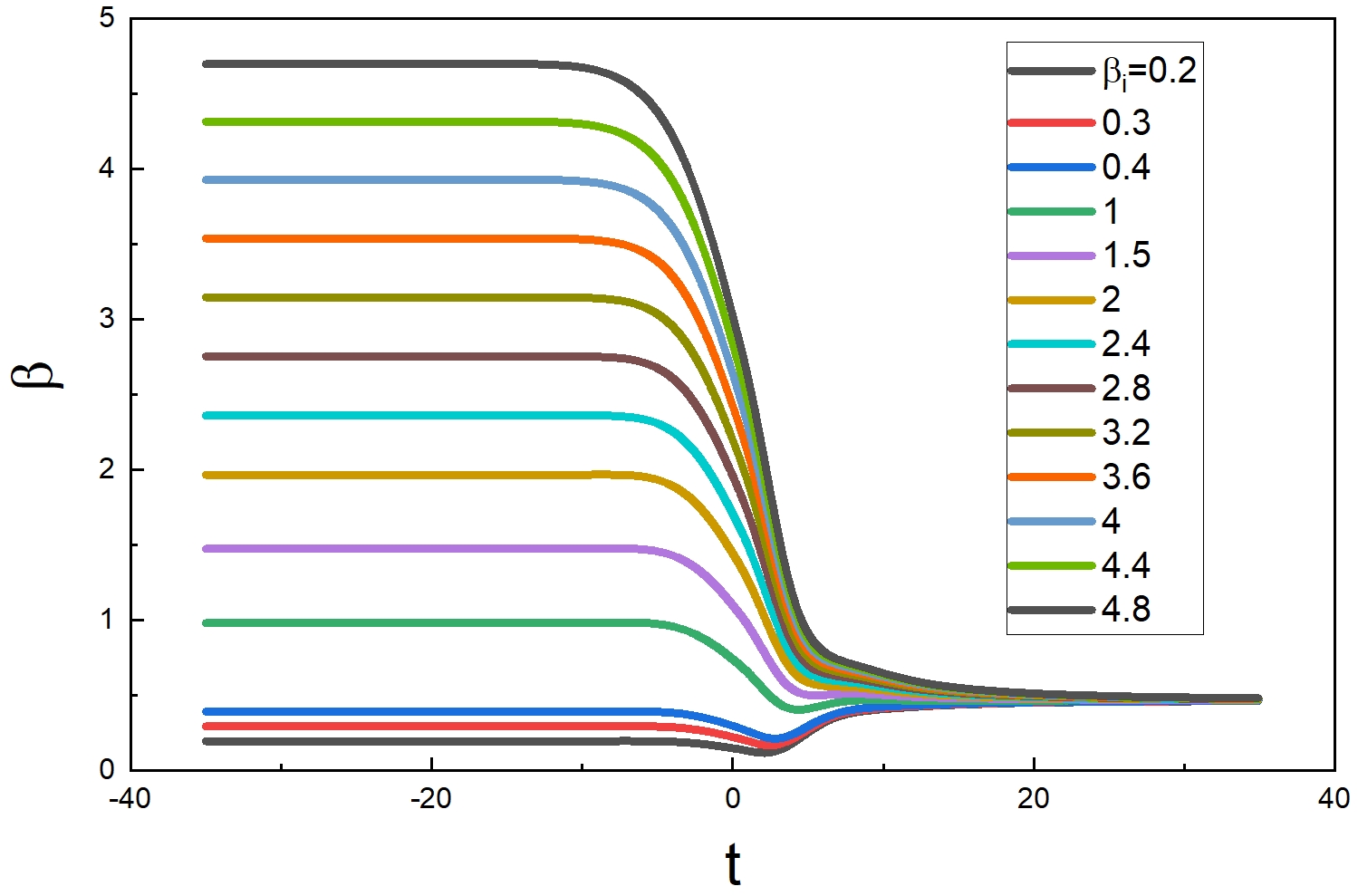}
    \includegraphics[width=0.45\textwidth]{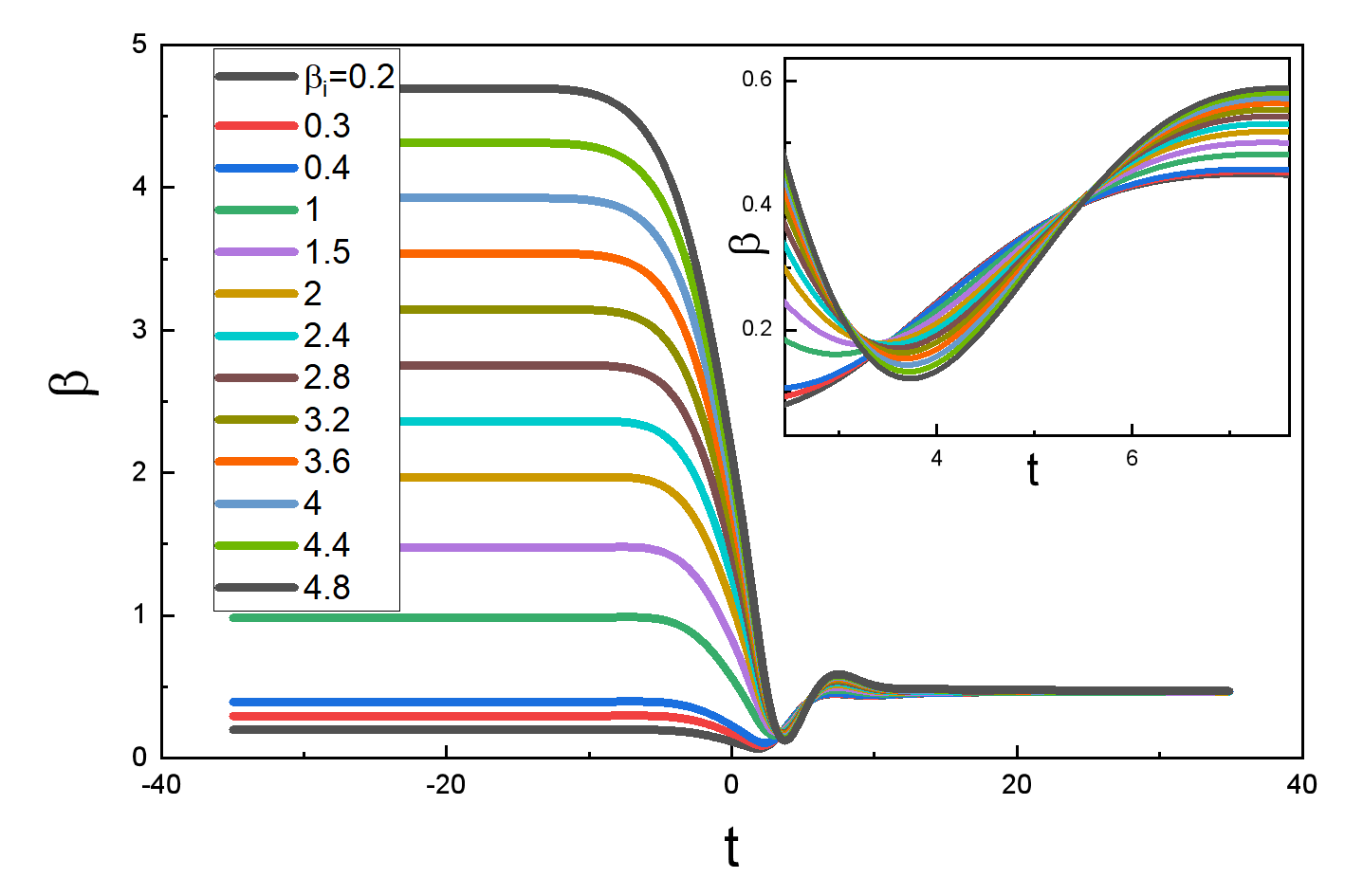}
    \includegraphics[width=0.45\textwidth]{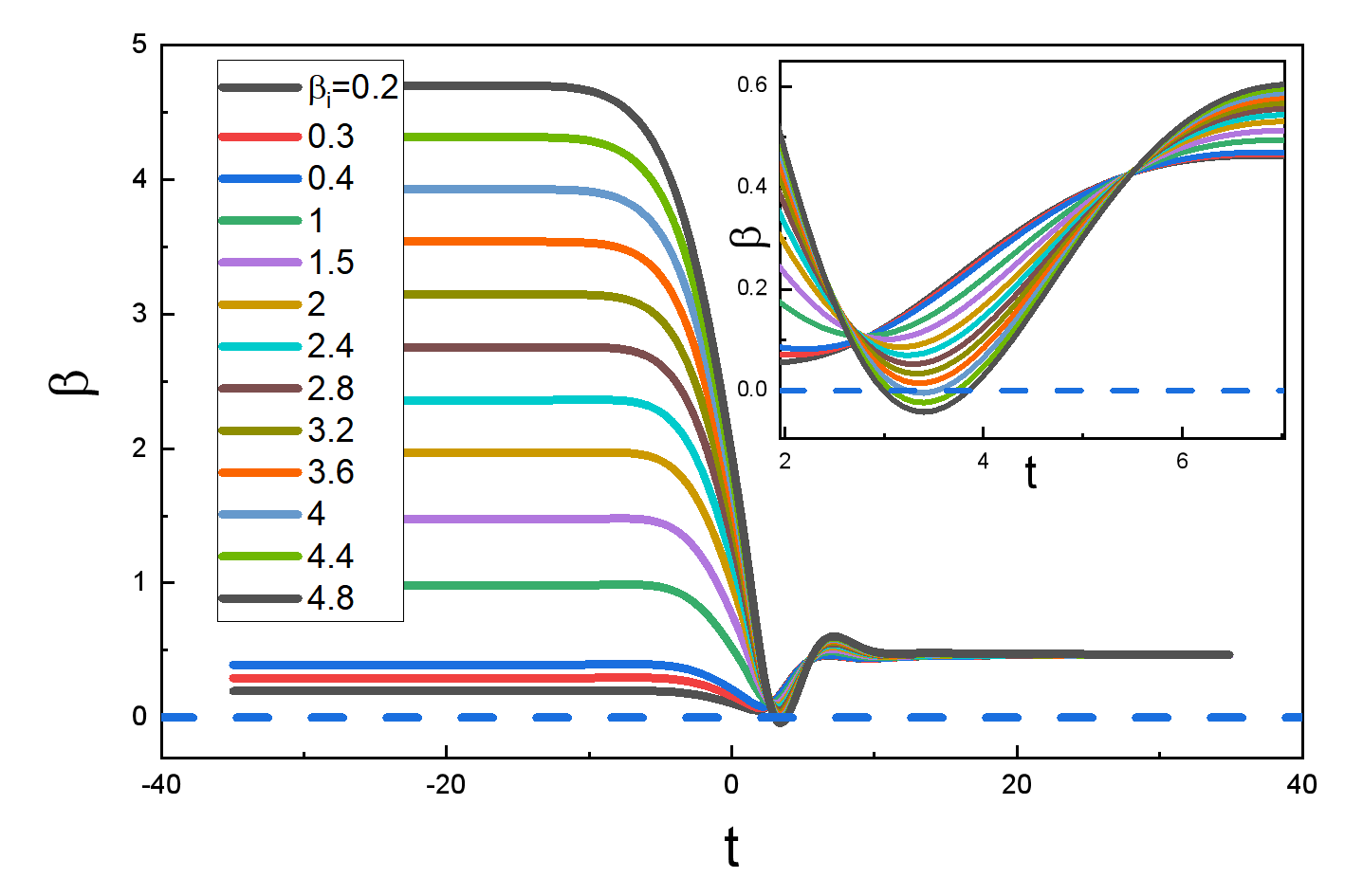}
    \caption{The evolution of inverse temperature $\beta$ vs time $t$. The coupling between the system and the bath is set to be (a) $V=0.4$, (b) $V=0.525$, and (c) $V=0.55$. For weak system-bath couplings, the thermalization processes behave as expected. The anomalous effects only exist for strong couplings. For all, $n=3,\, J=0.5,\, \beta_{\mathrm{bath}}=0.5.$}
    \label{fig:2}
\end{figure}

A solution to the KB equation for the Green's function $G^>_{\chi}(t,t')$ is shown in Fig.~\ref{fig:grn1}, where we have set the initial temperature of the system to be $\beta_i=2.4$ and the temperature of the thermal bath to be $\beta_{\mathrm{bath}}=0.5$. As in the equilibrium situation, the Green's function $G^>_{\chi}(t,t')$ decays rapidly away from the region near the diagonal slice $t=t'$. The system is coupled to the thermal bath at $t=0$ and is in equilibrium before that. For strong system-bath couplings, the Green's function exhibits rapid variations near the $t,t'=0$ and quickly relaxes to the steady state. 

\begin{figure}
    \centering
    \includegraphics[width=0.45\textwidth]{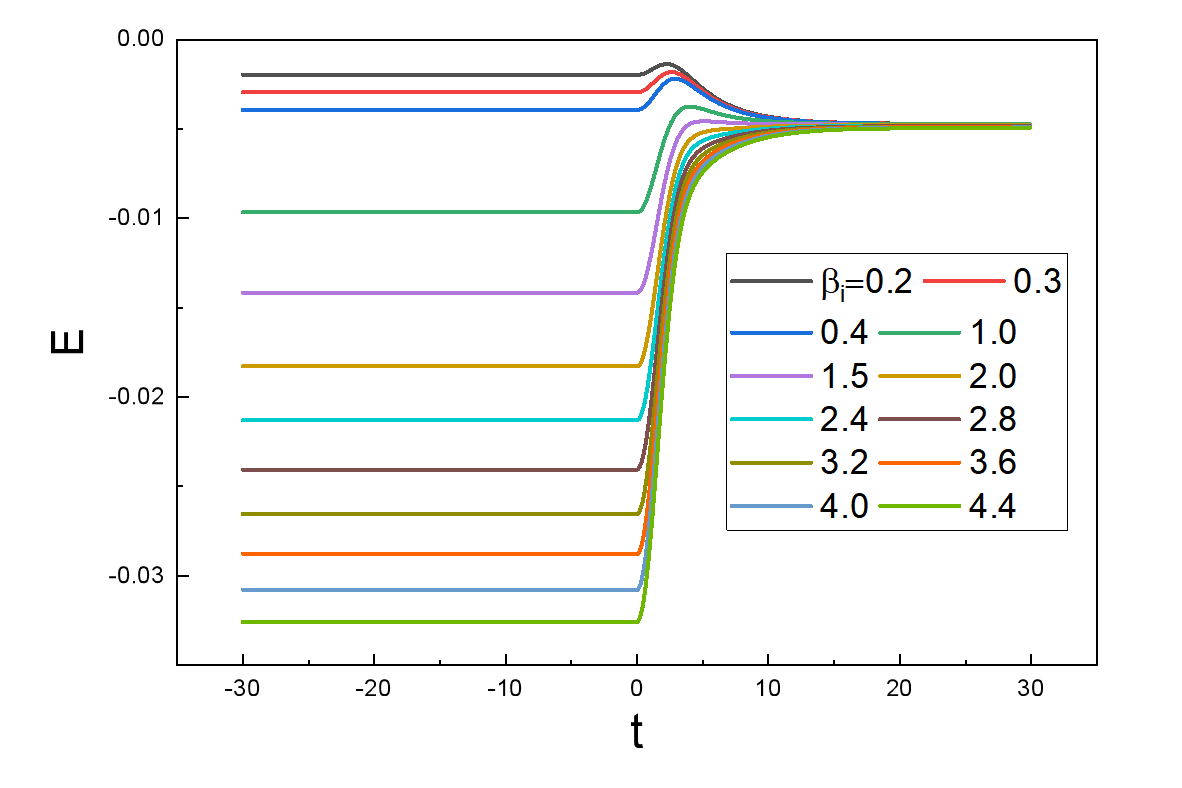}
    \caption{The time evolution of the total energy of the SYK model coupled with a bath. The parameters used are $n=3,\,\beta=0.5,\, V=0.525,\, J=0.5$.}
    \label{fig:et}
\end{figure}

In Fig.~\ref{fig:2} (a) and (b), we show that the Mpemba crossings only emerge when the coupling between the SYK system and the bath exceeds certain threshold value. For the weak couplings, the system's cooling dynamics resemble that of a quasi-equilibrium state so that our intuition from the equilibrium statistics works approximately in this situation. When the coupling is tuned up, the temperatures decay in the manner beyond the quasi-equilibrium description, the system with a lower (higher) initial temperature can be heated (cooled) to a temperature higher (lower) than the system starting from a higher (lower) initial temperature. The ``Mpemba crossing" emerges as the result of the strong out-of-equilibrium of the system induced by the strong coupling between the bath and the system. The strong coupling can also induce anomalous phenomena similar to overheating and overcooling caused by the energy input into the system by the system-bath coupling and the collective oscillation of the system [see Fig.~\ref{fig:et}]. The effective temperature can temporarily drops (rises) to a temperature lower (higher) than the asymptotic temperature in accordance with the ambient bath. In particular, when the coupling is sufficiently strong, the system is beyond the traditional equilibrium characterization and is accompanied by the emergence of negative temperatures according to the best fit of FDT [see Fig.~\ref{fig:2}(c)]. The negative temperature is usually considered as an indication that the system can dissipate heat to all equilibrium systems including those with $T=\infty$, which corresponds to the equal distribution of all energy levels. This appearance of the negative effective temperature is a distinct feature that the system is in strong nonequilibrium condition.

We remind that this oscillation of effective temperature is not necessarily just the redistribution of energy on different energy levels, the total energy can also manifest similar overheated phase due to the energy input induced by the strong coupling with the bath at $t=0$. The total energy can be expressed as \cite{zhang2019evaporation,bhattacharya2019quantum}:
\begin{align}
    E(t)=\sum_{i_1i_2i_3i_4}\frac{J_{i_1i_2i_3i_4}}{4!} \overline{\langle\chi_{i_1}\chi_{i_2}\chi_{i_3}\chi_{i_4}\rangle}=\frac{iJ^2}{4} \int_{-\infty}^t dt' \left(G^>(t,t')^4-G^<(t,t')^4 \right)\,.
\end{align}
An example of the dynamics of the total energy is shown in Fig.~\ref{fig:et}. Interestingly, anomalous oscillations and an overheated phase also manifest during the relaxation processes in systems strongly coupled to thermal baths, especially when system's initial temperature is not significantly higher than the bath temperature. However, no MPCs are identified emerge in the dynamics of the total energy. This suggests that the temperature dynamics are the result of the contributions both from the total energy as well as the nonequilibrium statistics. 

\begin{figure}
    \centering
    \includegraphics[width=0.43\textwidth]{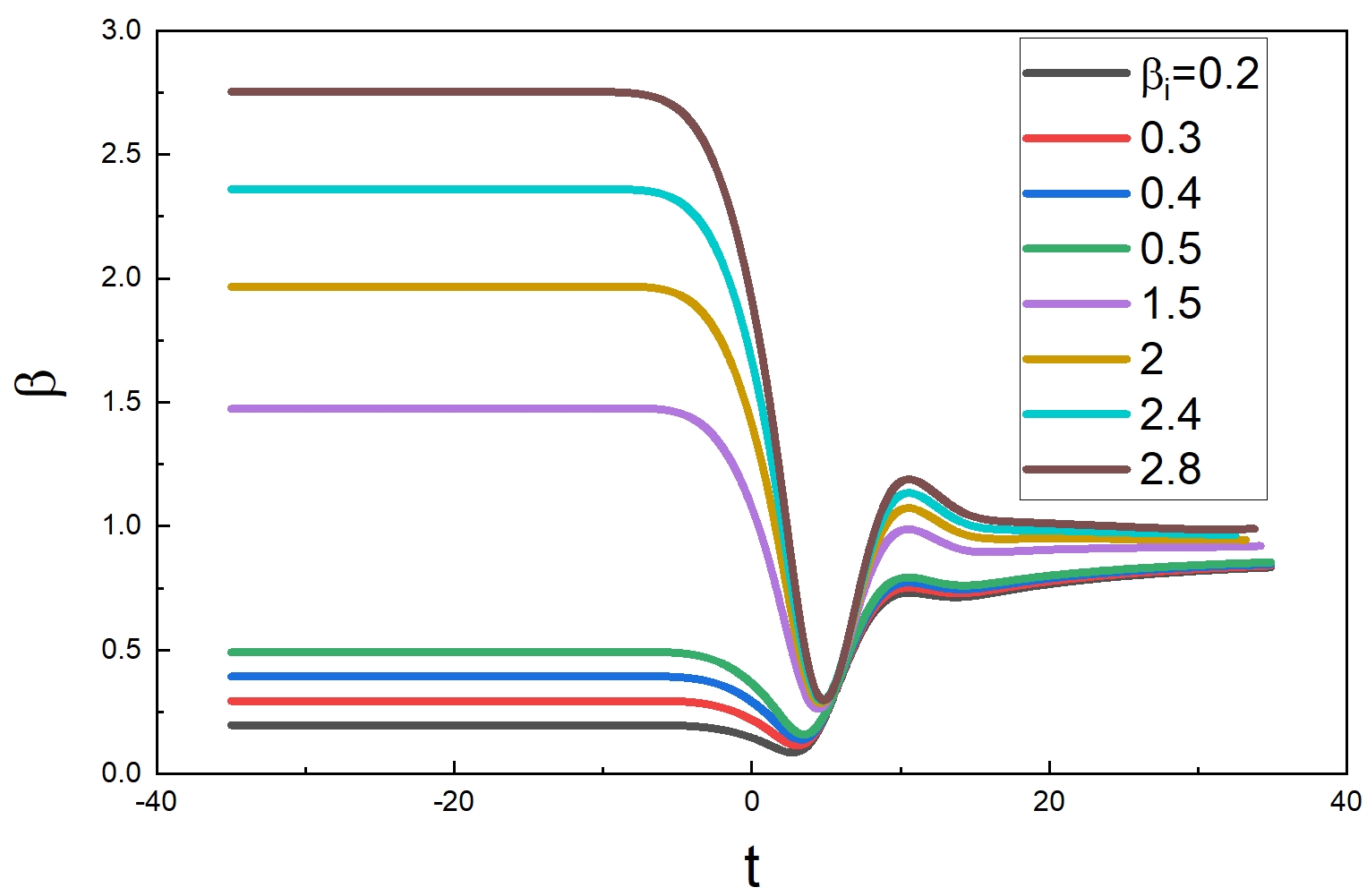}
    \includegraphics[width=0.43\textwidth]{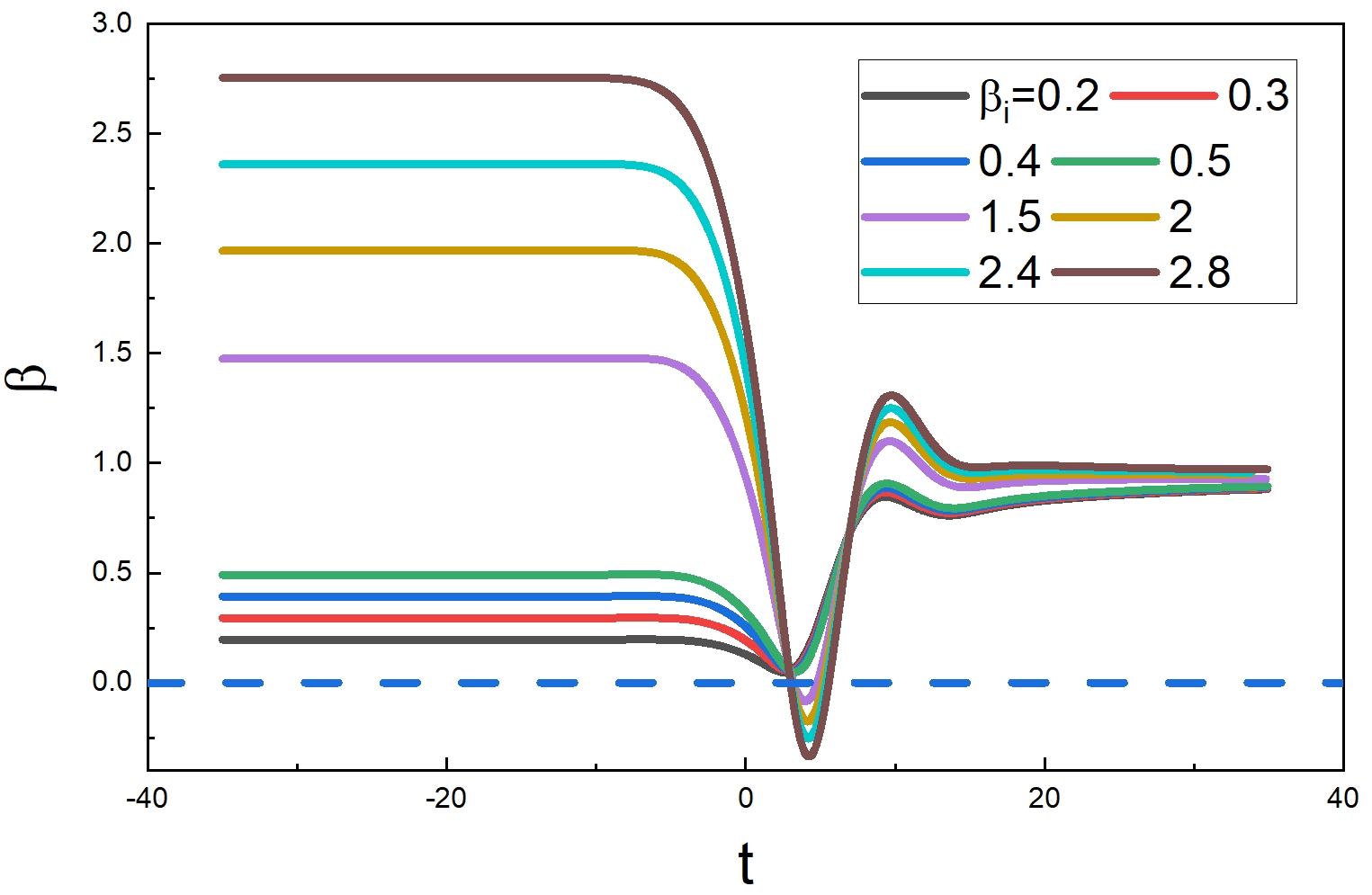}
    \caption{MPCs and negative temperatures emerge at sufficiently strong couplings for $n=1$ case. (a) The parameters are: $n=1,\,V=0.15$. (b) $n=1,\,V=0.175.$ For both (a) and (b), $J=0.5,\, \beta_{\mathrm{bath}}=0.5.$}
    \label{fig:negativet}
\end{figure}

For the SYK model with $n=1$, the oscillatory behavior after the quench is much more transparent than the $n=3$ case as shown in Fig.~\ref{fig:negativet}. The effective temperature is driven up dramatically due to the energy input impulse when the coupling is turned on at $t=0$ regardless of whether the bath temperature is higher or lower than the initial temperature of the SYK system. Another most noticeable feature from the simulation is that for $n=1$ model, a lower coupling strength is required for the emergence of the Mpemba crossing as well as the negative temperature. 

\begin{figure}[h!]
    \centering
    \includegraphics[width=1.0\columnwidth]{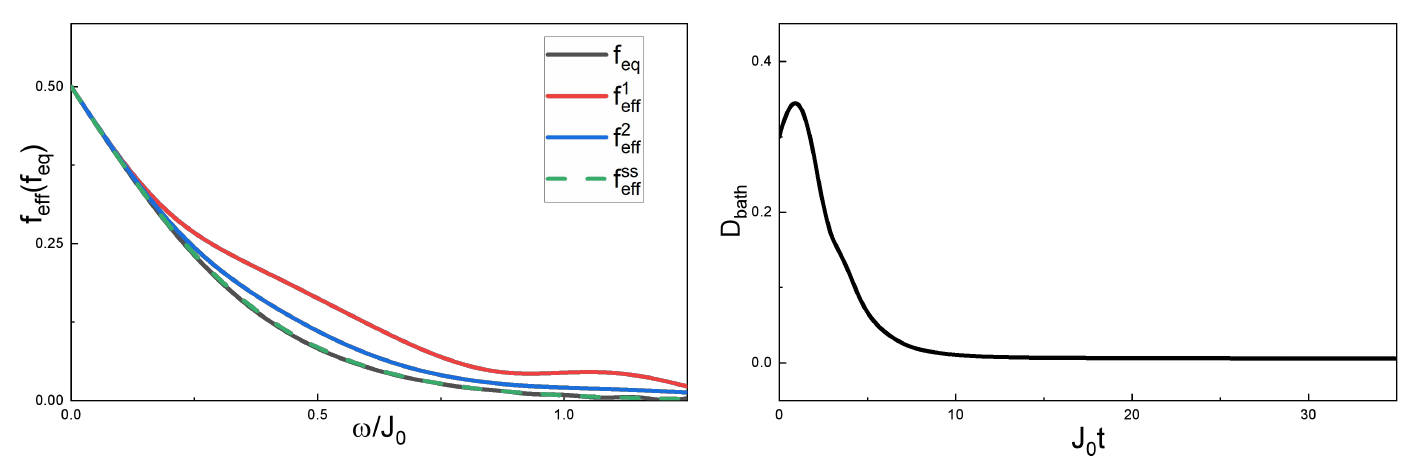}
    \includegraphics[width=0.5\columnwidth]{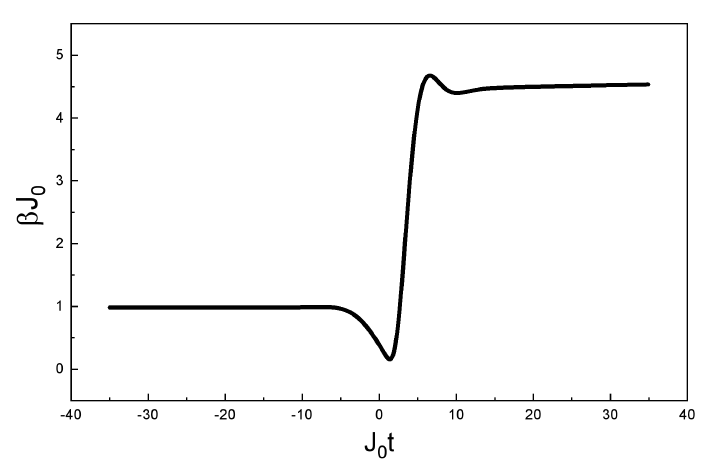}
    \caption{Nonequilibrium statistics in cooling process. (a) Energy distribution function when effective temperature reaches the bath temperature. (b) Distance function from equilibrium state. (c) The temperature trajectory for $\beta_{\rm initial}=1.0$. For all, $n=3,\, J=0.5,\,V=0.7,\,\beta_{\rm initial}=1.0,\, \beta_{\mathrm{bath}}=4.8.$}
    \label{fig:distribution48}
\end{figure}

In Fig.~\ref{fig:distribution48}, we illustrate the system statistics in the cooling process. Compared to heating, in a cooling process, the effective temperature will increase when the coupling is suddenly turned on, which results in the state further away from the steady state lying at a lower temperature. This feature is captured by the distance function shown in Fig.~\ref{fig:distribution48}(b). Similar to heating processes, the distribution function generally deviates from the equilibrium Fermi-Dirac distribution at higher frequencies and approaches the steady state distribution at late times.

\begin{figure}
    \centering
    \includegraphics[width=0.43\textwidth]{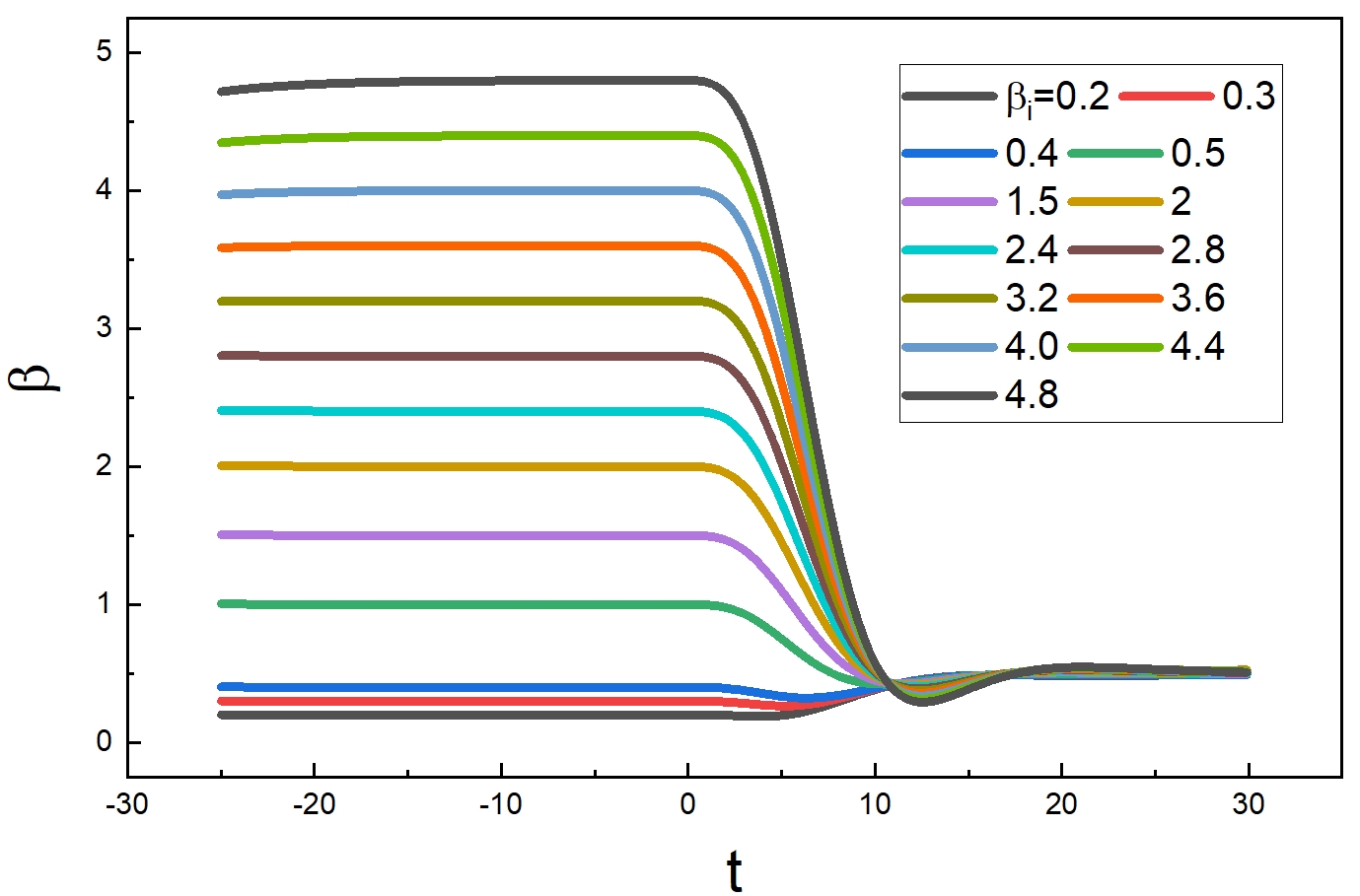}
    \includegraphics[width=0.43\textwidth]{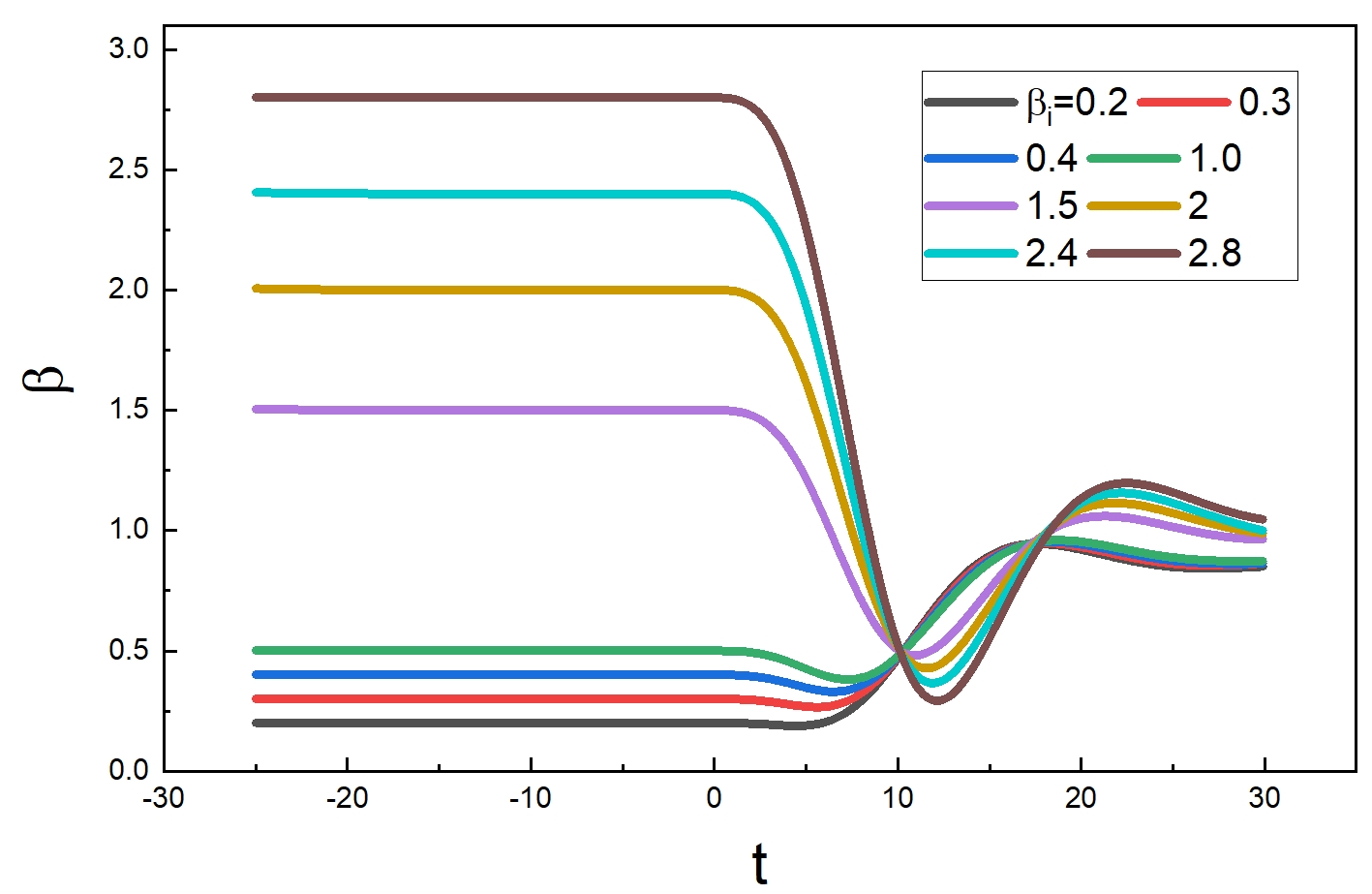}
    \caption{Dynamics of the inverse effective temperature according to Eq.~\eqref{eqbeta2}. (a) $n=3,\,\beta=0.5,\, V=0.4,\, J=0.5$. (b) $n=1,\,\beta=1.0,\, V=0.15,\, J=0.5$.}
    \label{fig:ap1}
\end{figure}

In summary, the quench dynamics of the SYK system coupled to a thermal bath reveal several anomalous behaviors, including oscillations in the effective temperature, temporal phases reminiscent of overheated and overcooled liquids, and Mpemba crossings between states that originate from different initial conditions. These observations hold true irrespective of the specific definition of the effective temperature. Results obtained using an alternative definition, Eq.\eqref{eqbeta2}, are presented in Fig.~\ref{fig:ap1}. Though the MPEs in this system only appear shortly after the quench instead of in the asymptotic states as seen in simple integrable quantum systems such as the quantum-dot and quantum-spin systems, it nonetheless serves as a proof of principle for the existence of dynamical anomalies during thermalization in chaotic systems. Further investigation covering a broader parameter space and with additional model variations is likely to strengthen what is shown in this study.


\section{MPCs in SYKs in contact with two different baths}
The SYK model coupled to a thermal bath, as described in the previous sections, can be easily generalized to the case with multiple baths. For example, we consider the SYK system simultaneously in contact with two thermal baths, the equations of motion is of the same KB form:
\begin{align}
 i\partial_{t_1}G^>_\chi(t_1,t_2)=\int d t_3 (&\Sigma^R_\chi(t_1,t_3)G^>_\chi(t_3,t_2)\notag \\ &+\Sigma^>_\chi(t_1,t_3)G^A_\chi(t_3,t_2)), \\
 -i\partial_{t_2}G^>_\chi(t_1,t_2)=\int d t_3 (&G^R_\chi(t_1,t_3)\Sigma^>_\chi(t_3,t_2)\notag \\ &+G^>_\chi(t_1,t_3)\Sigma^A_\chi(t_3,t_2)),
 \end{align}
 while the corrections to the self-energy in Eq.~\eqref{Eq:sigmas} have extra contributions from both thermal baths:
\begin{align}
\Sigma_\chi^{>}(t,t') &= -J^2 (G^>_{\chi}(t,t'))^3-V^2(-1)^{\frac{n+1}{2}}\theta(t)\theta(t')(G^>_{\psi}(t,t'))^n-V'^2(-1)^{\frac{n+1}{2}}\theta(t)\theta(t')(G^>_{\psi'}(t,t'))^n\,,\\
\Sigma_\chi^{<}(t,t') &= -J^2 (G^<_{\chi}(t,t'))^3-V^2(-1)^{\frac{n+1}{2}}\theta(t)\theta(t')(G^<_{\psi}(t,t'))^n-V'^2(-1)^{\frac{n+1}{2}}\theta(t)\theta(t')(G^<_{\psi'}(t,t'))^n\,,\\
\Sigma_\psi^{<}(t,t')&=-J^2 (G^<_{\psi}(t,t'))^3\,,\\
\Sigma_\psi^{>}(t,t')&=-J^2 (G^>_{\psi}(t,t'))^3\,.
\end{align}

\begin{figure}
    \centering
    \includegraphics[width=0.83\textwidth]{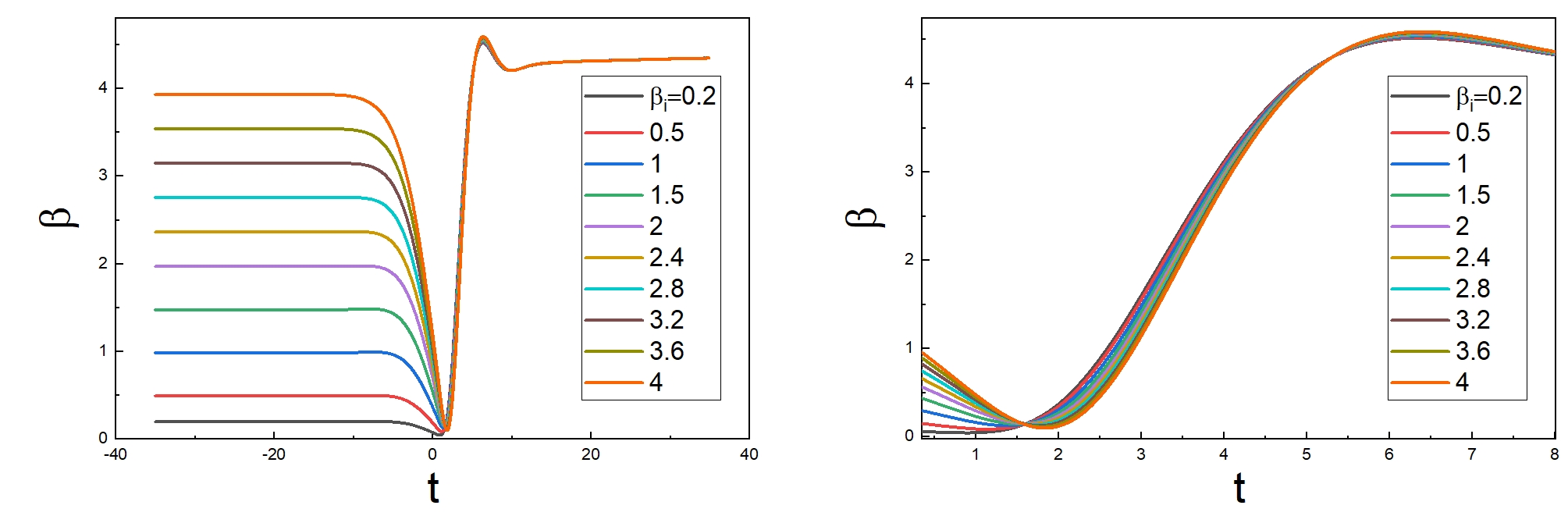}
    \caption{Mpemba crossings in the case of two different baths. (a) The full dynamics from equilibrium states to the steady states. (b) A zoomed-in view of the dynamics. The paramters used in the numerical simulations are $n=3,\, V_1=V_2=0.5,\, \beta_{bath1}=4.4,\, \beta_{bath2}=4.8$.}
    \label{fig:2bath}
\end{figure}

\begin{figure}
    \centering
    \includegraphics[width=0.4\textwidth]{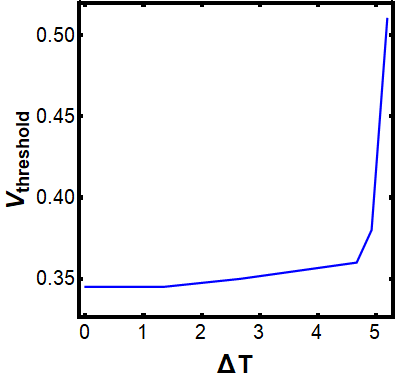}
    \caption{The threshold coupling $V_\mathrm{threshold}$ for the emergence of MPCs against the temperature bias between the two baths. The parameters used in the numerical simulations are: $n=3,\, J=0.5,\, \beta_{\mathrm{bath}}=0.5.$}
    \label{fig:2bath_vth}
\end{figure}

Then, we follow the procedure introduced in the previous sections and numerically solve Eqs.~\eqref{eq1} and \eqref{eq2} with the above self-energies. In contrast to the quantum dot system where MPE becomes more ubiquitous and emerges in less stringent criteria, in the open SYK system, larger coupling with the thermal bath is required to induce the Mpemba crossing. First of all, as shown in Fig.~\ref{fig:2bath}, the crossings of the different initial states start to emerge when the coupling is gradually tuned up. However, we notice that the crossings always occur $2N$ times, $N \in \mathbb{N}$, so that the MPE does not exist in the asymptotic final states, which requires $2N+1$ crossings. In contrast with the conventional MPE where temperature changes are monotonic and state crossings emerge at most once, here the temperature dynamics are more complicated and the relative positions of the effective temperatures oscillate. In addition, different from the nonequilibrium MPEs in the integrable systems such as the quantum dot system and double-fermionic system, systems with larger biases between two bath temperatures require stronger bath-system couplings to necessitate the emergence of the MPCs [see Fig.~\ref{fig:2bath_vth}]. Therefore, the temperature bias raises rather than lowers the threshold coupling, making it more different to observe the MPCs. This is another evidence that the MPCs in quantum integrable models have a different underlying mechanism from that in the SYK models. In Fig.~\ref{fig:2bath_vth}], the mean temperature of the two baths are fixed at around $\bar{T} \approx 5.3J$ while we vary the temperature bias between them. Therefore, the nonequilibrium enhancement of the MPE, as discovered in Ref.~\cite{wang2024mpemba}, only emerges in the integrable models where the MPEs are induced by peculiarity of initial conditions and does not apply to the chaotic SYK systems where MPEs are induced by nonequilibrium.

\section{Disappearance of MPCs in Liouvillian SYK}
Open SYK models can be modeled by the Lindblad master equations with linear jump operators. The Lindbladian SYK model is dual to the two-coupled SYK model where the real time $t$ in the Lindbladian SYK model plays the role of the inverse temperature in double SYK model \cite{garcia2023keldysh}. The common lore of the Lindblad equation is that the formalism gives a rough description of open systems while the very detailed information on the dynamics are erased. It is interesting to compare the quench dynamics derived from the SYK Lindbladian and that from the exact computation in the previous sections, and to find out whether the anomalies found in the previous sections can remain in this formalism. To compute the nonequilibrium Green's function, we employ the Schwinger-Keldysh formalism in the doubling Hilbert space with the vectorization of Liouvillian. In this way, the Lindblad equations of motion can be written in reminiscence of Schr\"odinger's equation and calculated from the Keldysh path integral. 

\subsection{Choi-Jamiolkwski isomorphism and Lindblad equation for SYK models}
To simplify the formalism for computing the Green's functions of the dissipative SYK model, we map the density matrix to a vector in the double Hilbert space via the Choi-Jamiolkwski (CJ) isomorphism \cite{zhou2023universal,zwolak2004mixed,zyxian24}, viz.,
\begin{align}
    \rho=\sum_{m,n} \rho_{m,n} |m\rangle\langle n| \qquad  \rightarrow \qquad |\psi^D_{\rho} (t)\rangle =\sum_{m,n}\rho_{m,n} |m\rangle \otimes |n\rangle  \in \mathcal{H}_L\otimes \mathcal{H}_R\,.
\end{align}
For states that satisfy the Lindblad equation:
\begin{align}
    \frac{d\rho}{dt}=-i[H,\rho]+2\gamma \sum_m L_m\rho L^\dag_m-\gamma\sum_m \{L_m^\dag L_m,\rho\}\,,
\end{align}
the state vector can be viewed or rewritten as the ``wavefunction" that satisfies the Schr\"odinger's equation:
\begin{align}
    i\partial_t \psi^D_{\rho} (t) =H^D \psi^D_{\rho} (t)\,,
    \label{eq:schrodinger}
\end{align}
where $H^D=H_s-iH_d$ is defined on the double space, $H_s=H_L\otimes\mathbb{I}_R-\mathbb{I}_L\otimes H_R^\mathrm{T}$, $ H_R^\mathrm{T}$ is the transpose of $H_R$, $H_L=H_R=H$, and $H_d$ is defined as follows:
\begin{align}
    H_d=\gamma \sum_m \left[-2 L_m \otimes L^{\dag \mathrm{T}}_m+ L^\dag L_m\otimes \mathbb{I}+\mathbb{I}\otimes (L_m^\dag L_m)^* \right]\,.
    \label{eq:H_d}
\end{align}
One can verify that the rule for a Lindbladian operator acting on the right Hilbert space (i.e., the operators to the right of the density matrix) is that it is simply changed to its transpose, viz., $L\rightarrow L^{\dag *}=L^\mathrm{T}$.

For the SYK model with the jump operators chosen to be $L_i=\sqrt{\mu}\psi_i$, the ground state of the Lindblad equation can be shown to be the maximally entangled state at the temperature $T=\infty$ \cite{garcia2023keldysh,zanoci2022energy}. The goal is to find a representation of the transpose of the Majorana fermion operator $\psi^\mathrm{T}$ using $\psi$. One algorithm to find such representation is provided. The Majorana operators are elements of the Cliford algebra $\mathcal{C}l(N)$ represented by the gamma matrices $\psi_k=\gamma_k$. For the double-site Majorana operators, one can choose the following representation for simplicity:
\begin{align}
    \psi_L^k=\gamma_k \otimes \mathbb{I}, \quad \psi_R^k=\gamma_c \otimes \gamma_k\,,
\end{align}
where $\gamma_c$ is the chiral matrix. We require that the transpose of the gamma matrix acting on the right Hilbert space is expressed as the right Majorana operator up to a constant and a global phase 
\begin{align}
    \mathbb{I}\otimes \gamma_k^\mathrm{T} \sum_j |j\rangle \otimes |j\rangle = \sum_j |j\rangle \otimes \gamma_k^\mathrm{T}|j\rangle = \alpha\, (U\otimes U) \sum_j \gamma_c |j\rangle \otimes \gamma_k |j\rangle
\end{align}
for some constant $\alpha$ and matrix $U$. We can rewrite this equation as
\begin{align}
   \sum_j |j\rangle \gamma_k^\mathrm{T}|j\rangle= \alpha \sum_j U \gamma_c |j\rangle U \gamma_k |j\rangle= \alpha \sum_j   |j\rangle U \gamma_k \gamma_c^{-1} U^{-1} |j\rangle\,.
\end{align}
Then, we have
\begin{align}
    \gamma_k^\mathrm{T}= \alpha \gamma_k \gamma_c^{-1} U^{-1}\,.
\end{align}
The equality of charge conjugation $C^{-1}\gamma_k C= \gamma_k^\mathrm{T}$ gives
\begin{align}
    U=C^{-1} e^{\frac{i \pi}{4}\gamma_c}, \quad \alpha = -i.
\end{align}
Therefore, from the above derivation we have the representation of the transpose of the Majorana operator
\begin{align}
     \mathbb{I}\otimes \psi_k^\mathrm{T}=-i \psi_R^k
\end{align}
With the above identification, one can write the Liouvillian operator in the doubled Hilbert space as 
\begin{align}
    \mathcal{L}=-i H^D= -i H_L^{\mathrm{SYK}}-i (-i)^q H_R^{\mathrm{SYK}}-i\mu \sum_i \psi_L^i \psi_R^i -\frac{\mu N}{2}
\end{align}
where $H^D$ is given in Eqs.~\eqref{eq:schrodinger} and \eqref{eq:H_d}. One can relate to the Keldysh formalism by introducing fields 
\begin{align}
   \psi^i(t^+)=\psi_L(t) \qquad \mathrm{and} \qquad \psi^i(t^-)=i\psi_R(t)
\end{align}  
living on $\mathcal{C}^+$ and $\mathcal{C}^-$, respectively. The extra negative sign in $\psi^i(t^-)=i\psi_R(t)$ is due to the inverse time direction from $+\infty$ to $-\infty$ on the $\mathcal{C}^-$ branch of the contour. The Keldysh closed-time contour is $\mathcal{C}=\mathcal{C}^+ \cup \mathcal{C}^-$. The Lindbladian SYK model in real time is dual to the Euclidean two-coupled SYK model by the following identifications \cite{garcia2023keldysh}:
\begin{align}
    &G_{LL}\rightarrow -i G_{LL},\qquad G_{RR}\rightarrow i G_{RR},\qquad G_{LR} \rightarrow G_{LR}, \qquad G_{RL}\rightarrow G_{RL}, \nonumber \\
    &\Sigma_{LL}\rightarrow i \Sigma_{LL},\qquad \Sigma_{RR}\rightarrow -i \Sigma_{RR},\qquad \Sigma_{LR}\rightarrow -\Sigma_{LR},\quad \Sigma_{RL}\rightarrow i \Sigma_{RL}\,.
\end{align}

\subsection{SYK Lindbladian}
Through Choi-Jamiolkwski isomorphism, the Lindbladian is mapped from the density matrix representation to the double-vector representation as follows \cite{sa2022lindbladian,kulkarni2022lindbladian}:
\begin{align}
  \mathcal{L}
  (\rho) = - i [H_{{\rm SYK}}, \rho]  +  \sum_{\alpha} \left[L^{\ }_{\alpha}\rho L^{\dag}_{\alpha}- \frac{1}{2} \{ L^{\dag}_{\alpha}L^{\ }_{\alpha}, \rho \} \right] \quad \rightarrow \quad \mathcal{L}=-i H^D= -i H_L^{\mathrm{SYK}}-i (-i)^q H_R^{\mathrm{SYK}}-i\mu \sum_i \psi_L^i \psi_R^i -\frac{\mu N}{2}
\end{align}
where $L^i=\sqrt{\mu} \psi^i$ is the linear jump operator and $H_{\text{SYK}}$ is given by
\begin{align}
H^{\text{SYK}}_{L(R)}[J_{i_1...i_q},\psi]=\sum_{i_1...i_q}\frac{J_{i_1...i_q}}{4!}\psi_{i_1,L(R)}...\psi_{i_q,L(R)}\,.
\end{align}
The partition function is written as 
\begin{align}
    Z=\int \mathcal{D}\psi_L \mathcal{D}\psi_R e^{iS[\mathcal{D}\psi_L,\mathcal{D}\psi_R]}\,,
\end{align}
where $L,\,R$ live in the doubled Hilber space $\mathcal{H}_L\otimes\mathcal{H}_R$ and the action is given by
\begin{gather}
    iS=\int_{-\infty}^{\infty} dt \left[-\frac{1}{2}\sum_i \psi^i_L \partial_t \psi_L^i-\frac{1}{2}\sum_i \psi^i_R \partial_t \psi_R^i +\mathcal{L} \right]\,.
    \label{eq:lindbladlagrangian}
\end{gather}
We can left out the constant term in the vectorized Liouvillian, viz.,
\begin{align}
     \mathcal{L}=-i H^D= -i H_L^{\mathrm{SYK}}-i (-i)^q H_R^{\mathrm{SYK}}-i\mu \sum_i \psi_L^i \psi_R^i \,.
\end{align}
By introducing the Keldysh contour $\mathcal{C}=\mathcal{C}^+\bigcup \mathcal{C}^-$ and the fields $\psi(t^+)=\psi_L(t)$ with $t^+\in \mathcal{C}^+$, $\psi(t^-)=i\psi_R(t)$ with $t^-\in \mathcal{C}^-$, one can rewrite the action as the following:
\begin{align}
      iS=-\int_{\mathcal{C}} dz \frac{1}{2}\sum_i^N \psi^i(z) \partial_z \psi^i(z)-i \int_{\mathcal{C}} dz i^{q/2}\sum_{i_1<\cdots<i_q}^N J_{i_1\cdots i_q}\psi^{i_1}(z)\cdots \psi^{i_q}(z)+\mu\int_{\mathcal{C}}dz dz' K(z,z')\sum_{i=1}^N \psi^i(z)\psi^i(z) \,,
\end{align}
with the dissipation kernel 
\begin{align}
    K(t_1^+,t_2^-)=\delta(t_1-t_2)\,,\quad   K(t_1^+,t_2^+)= K(t_1^-,t_2^+)= K(t_1^-,t_2^-)=0\,.
\end{align}
On the Keldysh contour, $dz=dt$ on $\mathcal{C}^+$ and $dz=-dt$ on $\mathcal{C}^-$. In terms of the collective variables $(G,\Sigma)$, where $\Sigma$ is the Lagrange multiplier of $G$, the action reads
\begin{gather}
    iS=\frac{N}{2} \{\mathrm{Tr}\log (i\partial_z-\Sigma)-\int_\mathcal{C}dzdz' \Sigma(z,z')G(z,z')-\frac{i^q J^2}{q}\int_\mathcal{C}dzdz'[G(z,z')]^q +2i\mu \int_\mathcal{C}dzdz'K(z,z')G(z,z')\}\,.
\end{gather}
We apply the following identities that for an operator $M$:
\begin{equation}
    \log (\det M) =\mathrm{Tr} \log M\,, \quad \delta[\det M]=\det M \mathrm{Tr}(M^{-1}\delta M)\,,
\end{equation}
and solve for the saddle point equation for the collective variables $\Sigma_{\alpha,\beta}(t_1,t_2)$ and $G_{\alpha,\beta}(t_1,t_2)$ by taking derivatives of the action with respect to $G_{\alpha,\beta}(t_1,t_2)$ and $\Sigma_{\alpha,\beta}(t_1,t_2)$, respectively. From the action, we can derive the saddle-point equations of motion for the Green's functions in the large-N limit, which are
\begin{equation}
    \begin{split}
       &(i\partial-\Sigma) G=\mathbf{1}_{\mathbb{C}},\\
       &  \Sigma_{\alpha\beta}(z,z') = -i^{q}J^2  G_{\alpha\beta}(z,z')^{q-1} + i\,\mu [K(z,z')-K(z',z)]\,.
\end{split}
\end{equation}
Restricting to the $\mathcal{C}^+$ and $\mathcal{C}^-$ contours and switching to the integral equation, we have the KB equation as follows:
\begin{equation}
    \begin{split}
       &i\alpha \partial_{t_1} G_{\alpha \beta}(t_1,t_2) - \int dt_3 \sum_{\gamma = +, -}\Sigma_{\alpha\gamma}(t_1,t_3) G_{\gamma \beta}(t_3,t_2) = \delta _{\alpha\beta}\delta(t_1-t_2)\,,\\
       &  \Sigma_{\alpha\beta}(t_1,t_2) = -i^{q}J^2 s_{\alpha\beta} G_{\alpha\beta}(t_1,t_2)^{q-1} + i\,\mu s_{\alpha\beta}\epsilon_{\alpha\beta} \delta(t_1-t_2)\theta(t_1)\theta(t_2)\,,
\end{split}  \label{eq:Liouvill1}
\end{equation}
where $s_{++}=s_{--}=1$, $s_{+-}=s_{-+}=-(-1)^{q/2}$, the Levi-CIvita symbols are $\epsilon_{+-}=1,\,\epsilon_{-+}=-1,\, \epsilon_{--}=\epsilon_{++}=0$, and $\alpha/\beta$ are ``$\pm$" signs. One can relate the lesser, time-ordered and anti-time-ordered Green's functions to the greater Green's function as follows:
\begin{align}
     &G^<(t_1,t_2)=- G^>(t_2,t_1)\,,\nonumber\\
     &G^{T}(t_1,t_2)=G^{++}(t_1,t_2)=\theta(t_1-t_2)G^{>}(t_1,t_2)+\theta(t_2-t_1)G^{<}(t_1,t_2)\,,\nonumber\\
     &G^{\tilde{T}}(t_1,t_2)=G^{--}(t_1,t_2)=\theta(t_2-t_1)G^{>}(t_1,t_2)+\theta(t_1-t_2)G^{<}(t_1,t_2)\,.
\end{align}
Combining the Kadanoff-Baym equation \eqref{eq:Liouvill1} and the symmetry relation
\begin{align}
    G^>(t_1,t_2)=\left(G^<(t_1,t_2)\right)^*=-\left(G^>(t_2,t_1)\right)^*\,,
\end{align}
we numerically solve for the dynamics of the Green's functions. 
Note that for the $-+$ component of the Eq.~\eqref{eq:Liouvill1}, the upper limit of the integration only extends to $\mathrm{max}(t_1,t_2)$ and the integrand vanishes for $t_3>\mathrm{max}(t_1,t_2)$.

\begin{figure}
    \centering
    \includegraphics[width=0.32\textwidth]{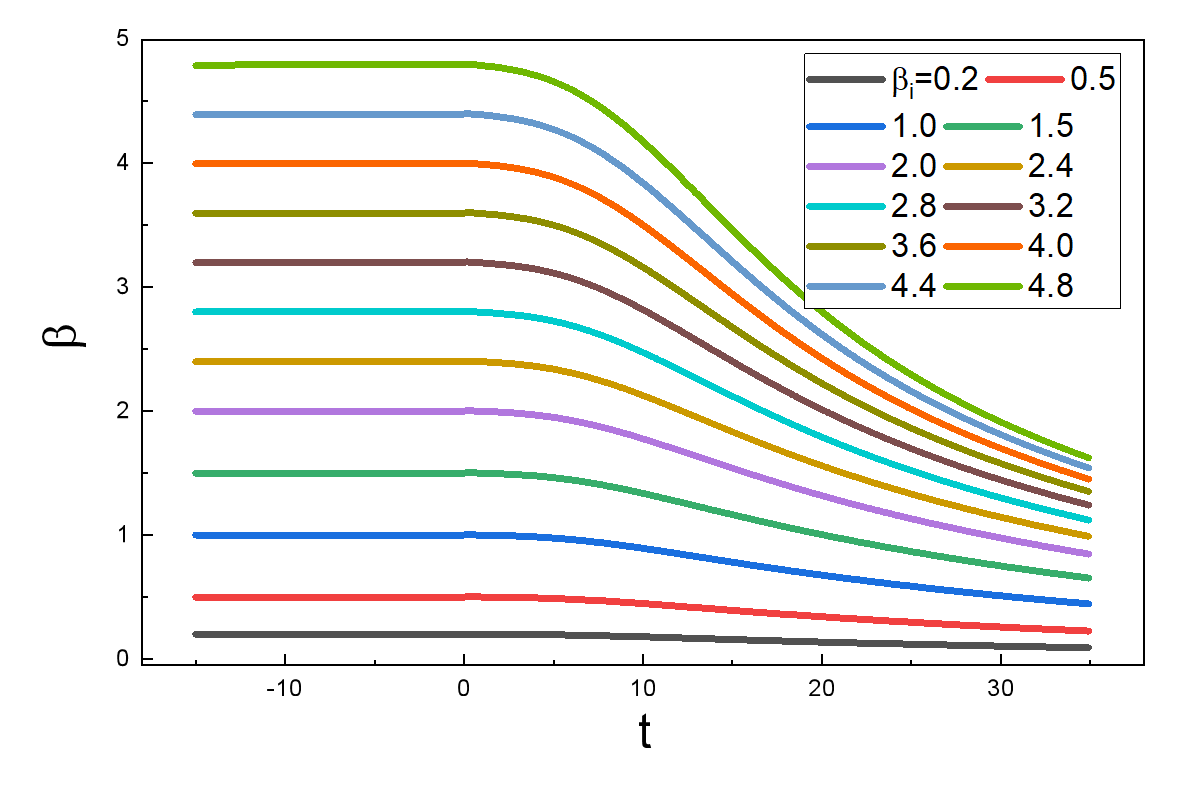}
    \includegraphics[width=0.32\textwidth]{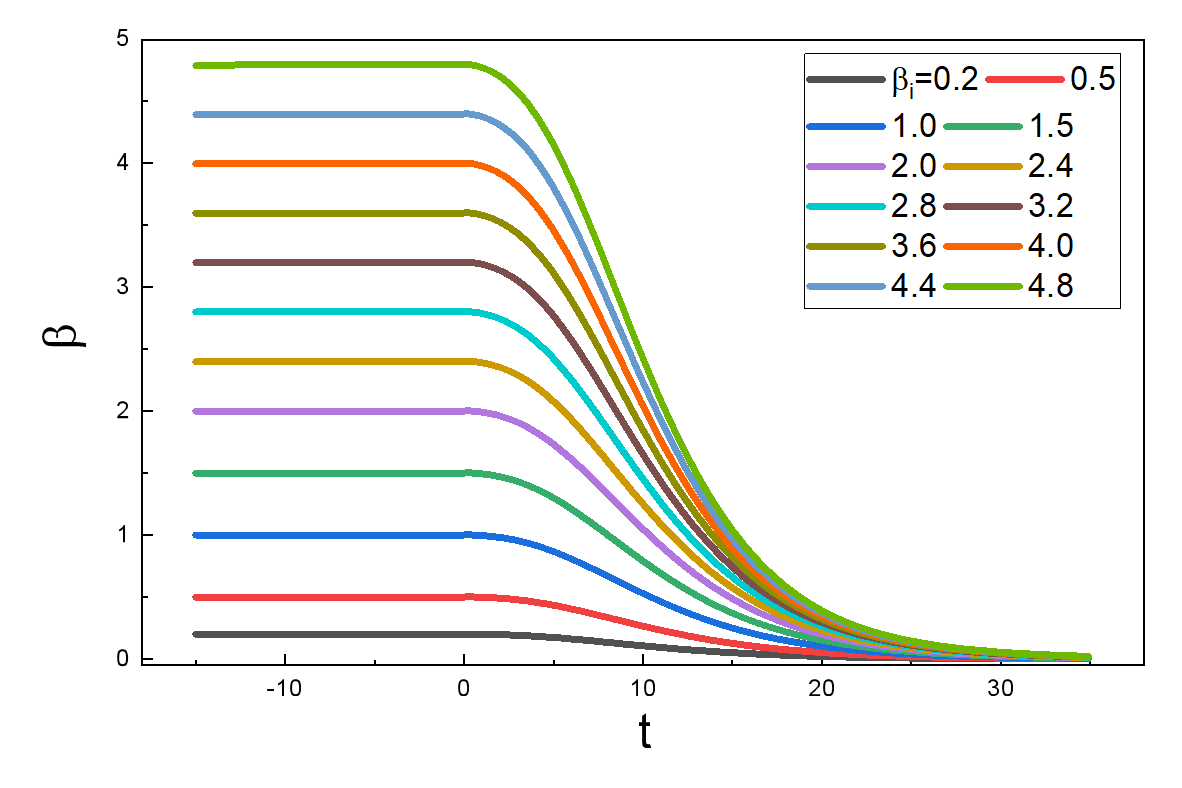}
    \includegraphics[width=0.32\textwidth]{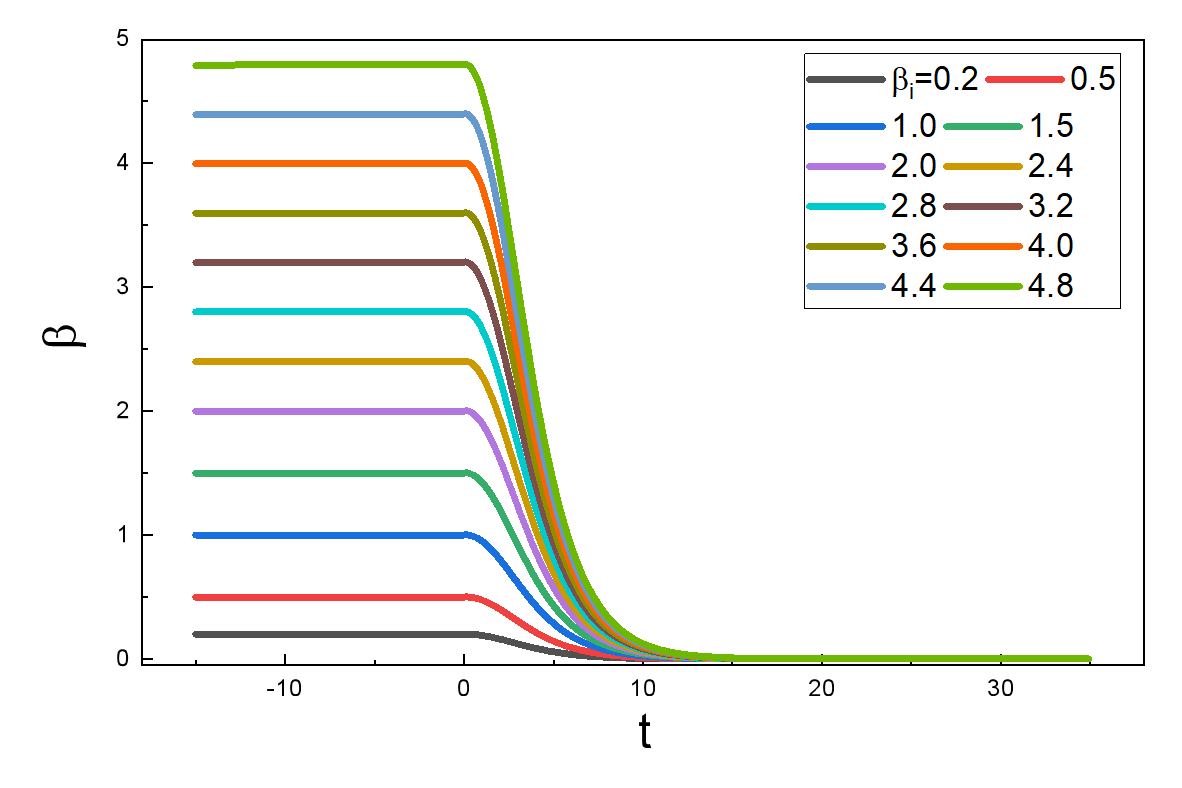}
    \caption{The effective temperature dynamics in the Lindblad description at the dissipative constant (a) $\mu=0.01$, (b) $\mu=0.05$, (c) $\mu=0.5$. The parameters used are $J=0.5,\, q=4$.}
    \label{fig:lind_t}
\end{figure}

\begin{figure}
    \centering
    \includegraphics[width=0.4\textwidth]{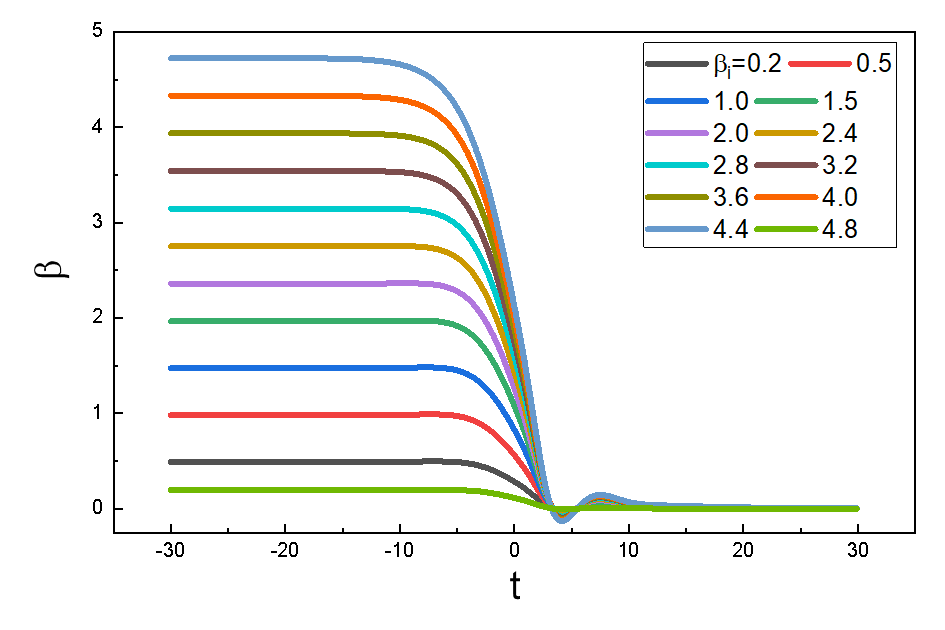}
    \includegraphics[width=0.4\textwidth]{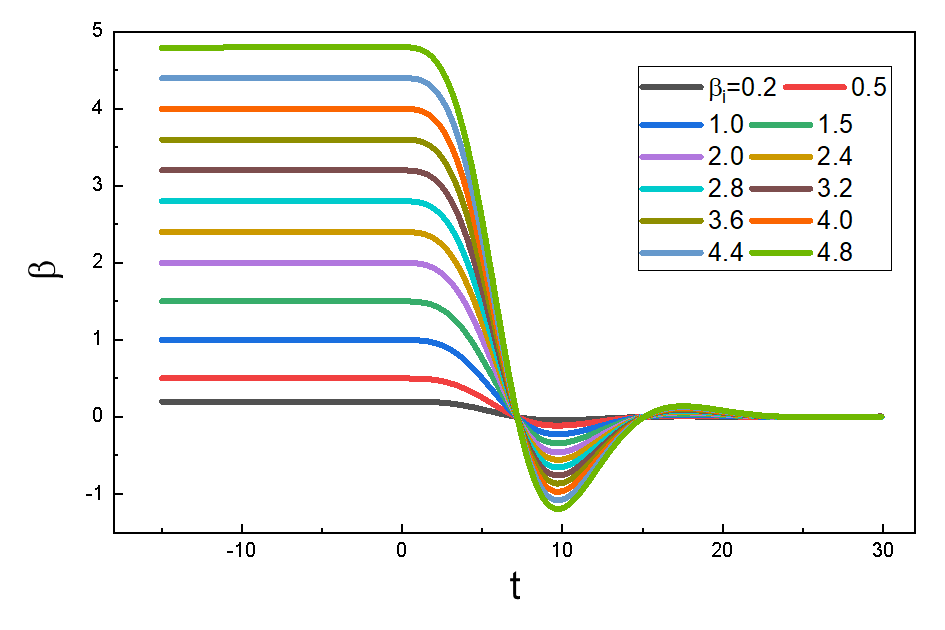}
    \caption{The effective temperature dynamics in the exact calculation of SYKs coupled with baths at infinite temperature. (a) The dynamics of the effective temperature according to definition Eq.~\eqref{eq:effectivetemp}. (b) The dynamics of the effective temperature according to definition Eq.~\eqref{eqbeta2}. The parameters used are $J=0.5,\, V=0.525,\, n=3$.}
    \label{fig:t_infty}
\end{figure}

\subsection{A different convention used for Liouvillian SYK model}
In this section, we provide a different set of definitions and the corresponding KB equations frequently used for the dissipative SYK model \cite{kulkarni2022lindbladian,kawabata2023dynamical}. The Green's functions are related as follows:
\begin{align}
G_{\alpha\beta}(t,t')=-i\left<\psi_\alpha(t)\psi_\beta(t')\right>=\begin{pmatrix}
G^{++}(t,t')&&G^{+-}(t,t')\vspace{3mm}\\ 
G^{-+}(t,t')&&G^{--}(t,t')
\end{pmatrix}=
\begin{pmatrix}
G^{T}(t,t')&&G^{<}(t,t')\vspace{3mm}\\ 
G^{>}(t,t')&&G^{\tilde{T}}(t,t')
\end{pmatrix}\,,
\end{align}
where $\alpha,\beta=\pm$, and $G^{T}(t,t')$ and $G^{\tilde{T}}(t,t')$ are defined as follows:
\begin{align}
    &G^{T}(t,t')=G^{++}(t,t')=i\left(\theta(t-t')G^{>}(t,t')+\theta(t'-t)G^{<}(t,t')\right)\,,\\
    &G^{\tilde{T}}(t,t')=G^{--}(t,t')=-i\left(\theta(t'-t)G^{>}(t,t')+\theta(t-t')G^{<}(t,t')\right)\,.
\end{align}
From the Lagrangian Eq.~\eqref{eq:lindbladlagrangian}, which is
\begin{align}
      iS=\int_{-\infty}^{\infty} dt -\frac{1}{2}\sum_i \psi^i_+ \partial_t \psi_+^i-\frac{1}{2}\sum_i \psi^i_- \partial_t \psi_-^i -i^{q+1} \sum_{i_1<\cdots <i_q}J_{i_1\cdots i_q} \psi^{i_1}_+\cdots \psi_+^{i_q} + i^{q+1} \sum_{i_1<\cdots<i_q}J_{i_1\cdots i_q}\psi^{i_1}_-\cdots \psi_-^{i_q}-i\mu \sum_i \psi_+^i(t) \psi_-^i(t)\,,
\end{align}
where we have left out the constant term proportional to $\mu N$. This action can be studied as the ordinary SYK in the large $N$ with the introduction of the collective fields $\Sigma$ and $G$. The action of the vectorized Lindblad equation in terms of the collective variables reads:
\begin{equation}
\begin{split}
     S[G,\Sigma]=&-\frac{iN}{2} \ln \mathrm{det}[-i(G_0^{-1}-\Sigma)]+\frac{i^{q+1}J^2N}{2q}\int_{t_1}^{t_2} dt_1dt_2 \sum_{\alpha,\beta}s_{\alpha,\beta}G_{\alpha,\beta}(t_1,t_2)^q\\
    &+\frac{iN}{2}\int_{t_1}^{t_2}dt_1dt_2\sum_{\alpha,\beta}\Sigma_{\alpha,\beta}(t_1,t_2)G_{\alpha,\beta}(t_1,t_2)-\frac{i\mu N}{2} \int_{t_1}^{t_2}dt [G_{+-}(t,t)-G_{-+}(t,t)]\,.
\end{split}
\end{equation}
We solve for the saddle point equation for the collective variables $\Sigma_{\alpha,\beta}(t_1,t_2)$ and $G_{\alpha,\beta}(t_1,t_2)$ by taking derivatives of the action with respect to $G_{\alpha,\beta}(t_1,t_2)$ and $\Sigma_{\alpha,\beta}(t_1,t_2)$, respectively. Then, we obtain the Kadanoff-Baym equation for the dissipative SYK model, which is
\begin{equation}
    \begin{split}
       &i\partial_{t_1} G_{\alpha \beta}(t_1,t_2) - \int dt_3 \sum_{\gamma = +, -}\Sigma_{\alpha\gamma}(t_1,t_3) G_{\gamma \beta}(t_3,t_2) = \delta _{\alpha\beta}\delta(t_1-t_2),\\
       &  \Sigma_{\alpha\beta}(t_1,t_2) = -i^{q}J^2 s_{\alpha\beta}G_{\alpha\beta}(t_1,t_2)^{q-1} + \theta(t_1)\theta(t_2)\mu\epsilon_{\alpha\beta} \delta(t_1-t_2).
\end{split}  \label{eq:KB Liouvill}
\end{equation}
Here $s_{++}=s_{--}=1$, $s_{+-}=s_{-+}=-(-1)^{q/2}$ and $\epsilon_{\alpha \beta}$ is the Levi-CIvita symbol defined as $\epsilon_{+-}=1,\,\epsilon_{-+}=-1,\, \epsilon_{--}=\epsilon_{++}=0$. The time derivative with respect to $t_2$ can be obtained by the symmetry relation $\partial_{t_2}G^>(t_1,t_2)=\left(\partial_{t_2}G^>(t_2,t_1)\right)^*$ or $\partial_{t_2}G^>(t_1,t_2)=-\partial_{t_2}G^<(t_2,t_1)$. For Majorana fermions, the differential equation with respect to $t_1$ suffices for solving the Wightman Green's function due to the symmetry of Green's function given by
\begin{align}
    G^>(t_1,t_2)=-\left(G^<(t_1,t_2)\right)^*=\left(G^>(t_2,t_1)\right)^*\,, \quad \mathrm{or}\quad G^>(t_1,t_2)=-G^<(t_2,t_1)\,.
\end{align}
Notice the sign difference from the two-coupled SYK models.

It is important to note the difference between the above definition and the notation used in the SYK systems coupling with a bath, especially when carrying out numerical simulations. In the above Liouvillian setup, the initial condition can be set to the thermal state at the inverse temperature $\beta$. The Green's functions of the initial thermal state in this formalism is related to the Green's function in the isolated SYK model by the following identifications: 
$G_l^{++}(t,t')=G_i^{++}(t,t')$, $G_l^{+-}(t,t')=-i G_i^{+-}(t,t')$, $G_l^{-+}(t,t')=-i G_i^{-+}(t,t')$, and $G_l^{--}(t,t')=- G_i^{--}(t,t')$, where $G_l^{\alpha,\beta}$ is the Green's function in the Liouvillian SYK and $G_i^{\alpha,\beta}$ is the Green's function for an isolated SYK model.
In this setsup, the effective temperature can be defined through FDT as:
\begin{gather}
    \beta(t)=\left.\frac{2 \cdot \mathrm{Re}(G_{K}(\omega,t))}{\omega \cdot \mathrm{Re}(G_{R}(\omega,t))}\right|_{\omega\rightarrow 0}\,.
\end{gather}

\begin{figure}
    \centering
    \includegraphics[width=0.4\textwidth]{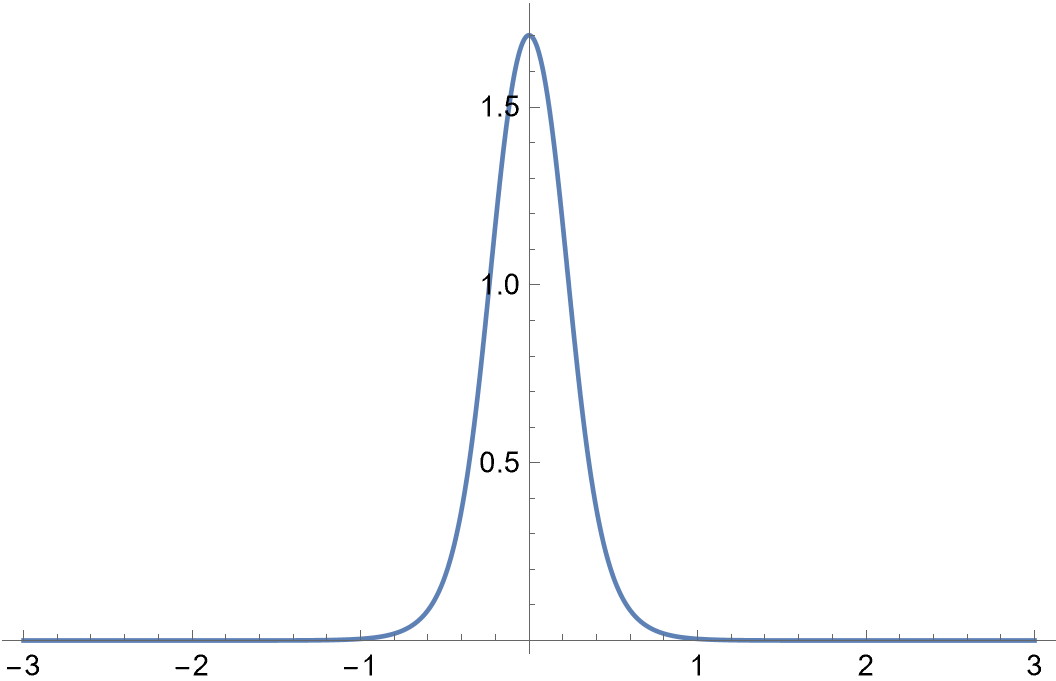}
    \includegraphics[width=0.4\textwidth]{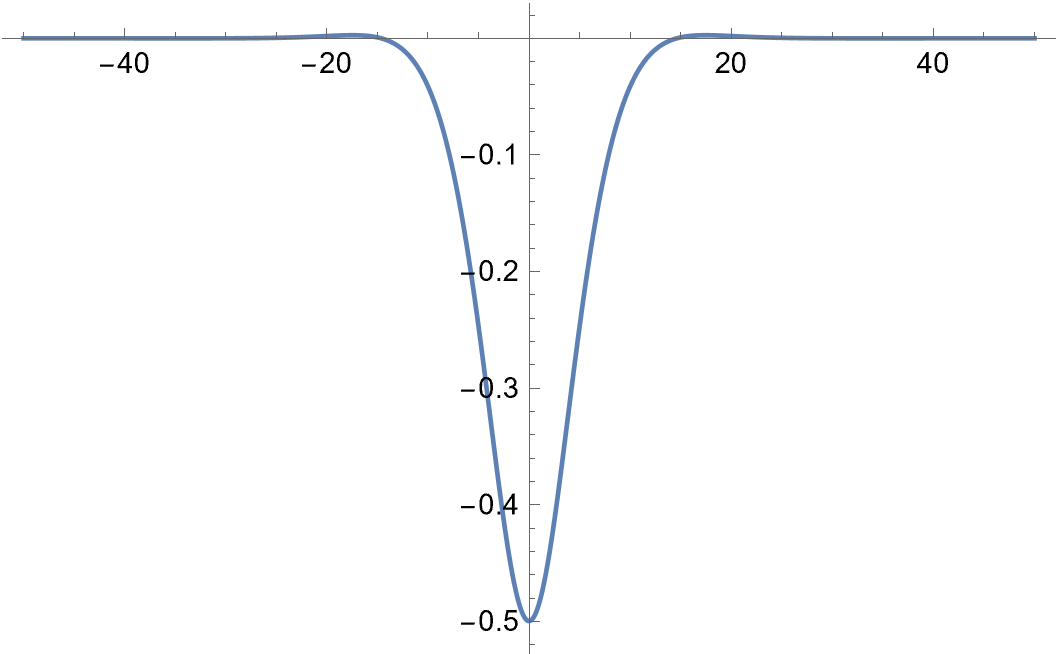}
    \caption{The imaginary part of the Green's function at infinite temperature from solving the self-consistency equation Eq.~\eqref{eq:consistencyeq}. (a) The spectrum of the Green's function Im($G(\omega)$) at $\beta\rightarrow 0$. (b) Im($G(t)$) at $\beta\rightarrow 0$. For a finite SYK coupling $J$, the imaginary part of the correlation function does not return to the delta function. The imaginary part of the correlation function returns to the delta function only in the strongly-coupled limit.}
\end{figure}

\subsection{Numerical results and comparisons}
To investigate quantum chaotic systems, we focus on SYK$_4$ in our numerical study. We vary the dissipative constant $\mu$ and the initial temperature of the system and numerically calculate the real-time dynamics of the SYKs dictated by the KB equation Eq.~\eqref{eq:KB Liouvill}. The initial condition of the correlation functions can be derived from the correlators in the previous section, viz., for $t,t'<0$: 
\begin{eqnarray}
     &G_l^{++}(t,t')=G_i^{++}(t,t')\,,\quad &G_l^{+-}(t,t')=-i G_i^{+-}(t,t')\,,\\
    &G_l^{-+}(t,t')=-i G_i^{-+}(t,t')\,,\quad &G_l^{--}(t,t')=- G_i^{--}(t,t')\,,
 \end{eqnarray}
where $G_i$ with subscript ``i" denotes the corresponding correlator defined in Eq.~\eqref{eq:grn}. We set the system's initial Green's function $G^>_i(t,t')=G^{-+}_i(t,t')$ before the quench and then use the Eq.~\eqref{eq:KB Liouvill} to generate the dynamics of the system.

Interestingly, we did not find any anomalies in the dynamics of the effective temperature as those emerged in the exact solutions of the SYKs coupled with SYK baths. As shown in Fig.~\ref{fig:lind_t}, the inverse of the effective temperature drops approximately exponentially at late times, corresponding to an exponential rise in temperature. The dynamics share resemblance with that in quasi-equilibrium approximation where the inverse temperature decreases smoothly and monotonically after $t=0$ with trajectories from different initial states well-separated before converging to the steady state. In contrast to the Lindbladian dynamics, we can also compute the inverse temperature dynamics by coupling the SYK model to an infinite-temperature SYK bath. The results are shown in Fig.~\ref{fig:t_infty}, where we identify similar collective oscillations of the system temperatures and MPCs in the heating process.

The two sets of numerical results show a clear difference between direct solving the KB equations with infinite-temperature thermal baths and solving the SYK Lindblad equation. It is worth mentioning that the two SYK models do not relax to the same equilibrium state, even though both final states have inverse temperature zero. One way to understand such distinction is by noticing that the KB equation of the Lindbladian SYK can be related to the KB equation of an SYK system coupled to a thermal bath by substituting the bath’s time correlation in the self energies with one proportional to a Dirac delta function in time, i.e., $i \delta(t-t')$ as in Eq.~\eqref{eq:Liouvill1}. In many weakly interacting quantum systems, the vanishing width of the Green's function typically corresponds to the system reaching an infinite temperature. However, the SYK model has nontrivial real-time dynamics even at the temperature $T=\infty$ and the spectral function remains approximately the unchanged beyond a certain temperature threshold. The fact that the SKY Green's function has finite width proportional to the SYK coupling was also pointed out in Ref.~\cite{zhang2021obstacle}, where it was argued that the quasi-particle decay rate satisfies $\Gamma \approx \frac{1}{\sqrt{q-1}2^{q/2-2}}J$ assuming the form of Green's function $G^R(\omega)\approx \frac{1}{\om+i\Gamma}$. This difference in the spectrum causes a clear distinction of the Lindbadian SYK model from a dissipative SYK model in a thermal bath.

To recapitulate, we find that the Lindbladian SYK approach, though mimicking SYK model in an infinite-temperature bath, does not admit the same MPCs as in the SYK models coupled to thermal baths at $\beta=0$. The results show that the common lore about the Lindblad equations--averaging over fast modes and suppressing quantum oscillations in the system--effectively remains valid in this context. One of the reasons for the difference is that SYK models still have nontrivial real-time dynamics at $T=\infty$ and finite widths in its imaginary part of the two-point function, contrasting the flat spectrum across all frequencies of the delta function in the Lindblad equation. 

\bibliography{bib}

\end{document}